\documentclass[twoside]{dis04}

\usepackage{cite}

\def\runauthor{W. H. Smith}
\def\shorttitle{Recent Results from ZEUS}

\def\perc{\,\%}

\newcommand{\ksp}{K^{0}_{S}\> p}
\newcommand{\kspb}{K^{0}_{S}\> \bar{p}}
\newcommand{\ksppb}{K^{0}_{S}\> p\>(\bar{p})}
\newcommand{\coll}{Collaboration}
\newcommand{\etal}{ {\it et al.,} }
\newcommand{\gev}{\; \mathrm{GeV}}
\newcommand{\mev}{\; \mathrm{MeV}}

\def\z0{Z^0}
\def\mz{M_Z}
\def\as{\alpha_s}
\def\oalphas2{{\cal O}(\alpha\as^2)}

\def\asz{\as(\mz)}
\def\asmz#1#2#3#4#5#6{\asz = #1\pm #2\ {\rm (stat.)}\ ^{+#4}_{-#3}\ {\rm (exp.)}\ ^{+#6}_{-#5}\ {\rm (th.)}}
\def\etaphi{\eta-\varphi}
\def\etjet{E_T^{\rm jet}}

\def\be{\begin{equation}}
\def\ee{\end{equation}}
\def\bea{\begin{eqnarray}}
\def\eea{\end{eqnarray}}

\begin{document}

\title{RECENT RESULTS FROM THE ZEUS EXPERIMENT}

\author{ Wesley H. Smith \\ \small{(on behalf of the ZEUS Collaboration)}}

\address{ Physics Dept., University of Wisconsin, 
Madison, WI 53706 USA \\ E-mail: wsmith@hep.wisc.edu}

\maketitle

\abstracts{A summary of recent results from ZEUS is presented. New ZEUS
results from HERA-1 data include Structure Functions, QCD fits, analysis
of hadronic final states, precision measurements of $\alpha_s$,
production of heavy flavor mesons and baryons and studies of
diffraction. Results from the new HERA-II running include the
measurement of the cross section for polarized charged current events
and charm events tagged with the new ZEUS vertex detector.}

\maketitle

\section{Introduction}

There are new ZEUS results on total neutral and charged current
differential cross sections, structure functions, and their QCD fits.
Analyses of the hadronic final states in these data have produced
several precision measurements of $\alpha_s$ and have been used to test
new NLO QCD calculations.

ZEUS data show evidence for pentaquark baryons with strangeness.
Precision measurements of charm production in DIS may now be used to
check NLO calculations and constrain the gluon content of the proton.
Measurements of B production in photoproduction and DIS are showing
improving agreement with NLO QCD. Electroproduction of $\phi$ and
J/$\psi$ mesons show consistency with VMD and Regge phenomenology as
well as with pQCD. Diffractive charm production can now begin
to discriminate amongst diffractive parton distribution functions.

Following the upgrade to HERA-II, the ZEUS detector turned on with new
microvertex and forward tracking detectors. In 2004, HERA achieved
longitudinal polarization of $ > 50\perc$ and has begun to deliver
luminosities approaching the required rate to reach the luminosity goal.
Physics results from the beginning of the HERA-II luminosity running
with $ 33\perc$ polarized positrons show that the ZEUS detector upgrades
are working well.

These ZEUS results are discussed below and presented in greater detail
in the individual contributions in these proceedings.

\section{Cross Sections, Structure Functions and QCD fits}

\begin{figure}
\begin{center}
\begin{minipage}[c]{0.48\textwidth}
\psfig{figure=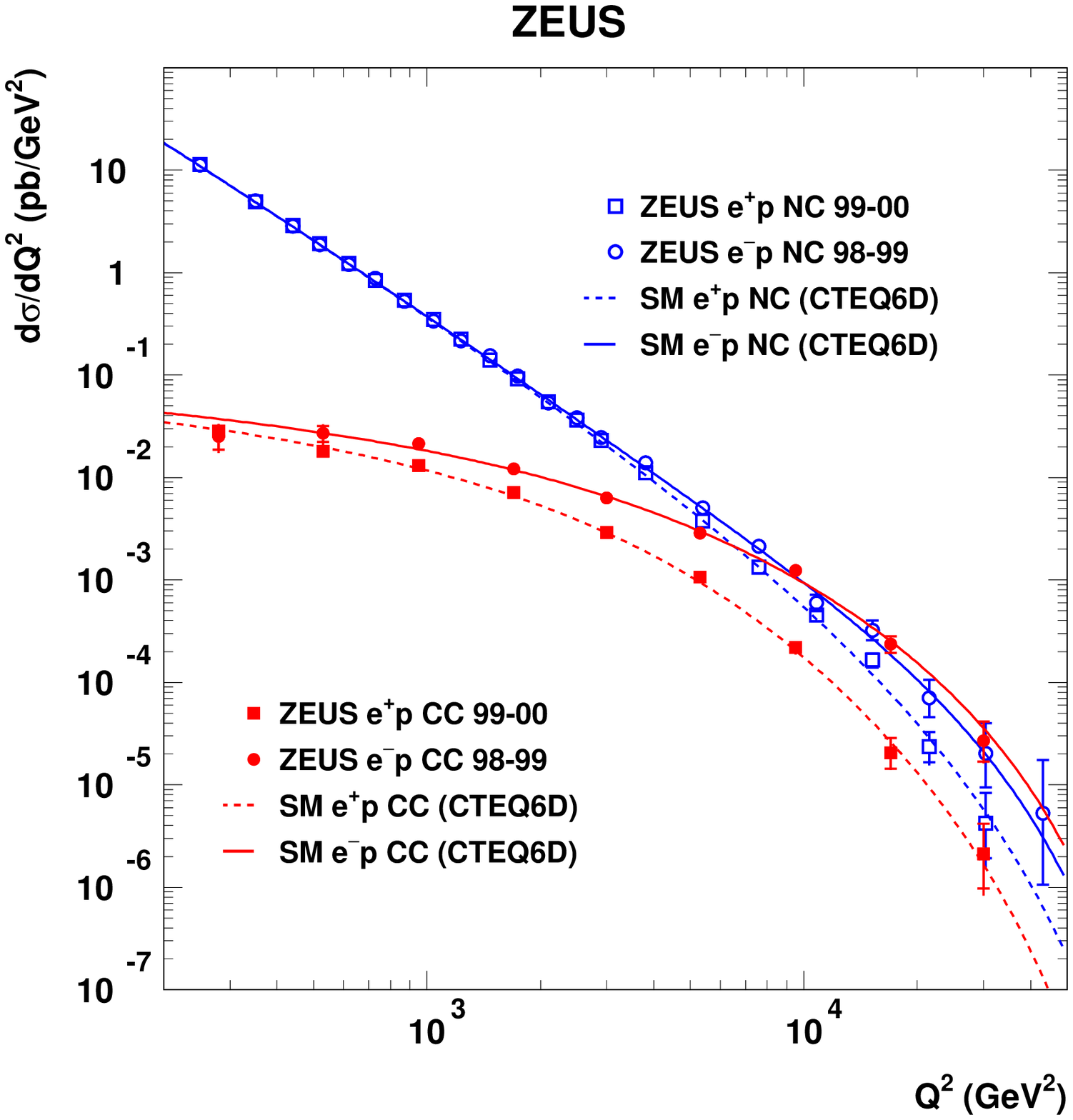,width={0.98\textwidth}}
\caption{The {NC}  and {CC} cross sections versus $Q^2$ for $e^+p$ and $e^-p$ 
  interactions.}
\label{fig:NCepem}
\end{minipage}
\hfill
\begin{minipage}[c]{0.48\textwidth}
\psfig{figure=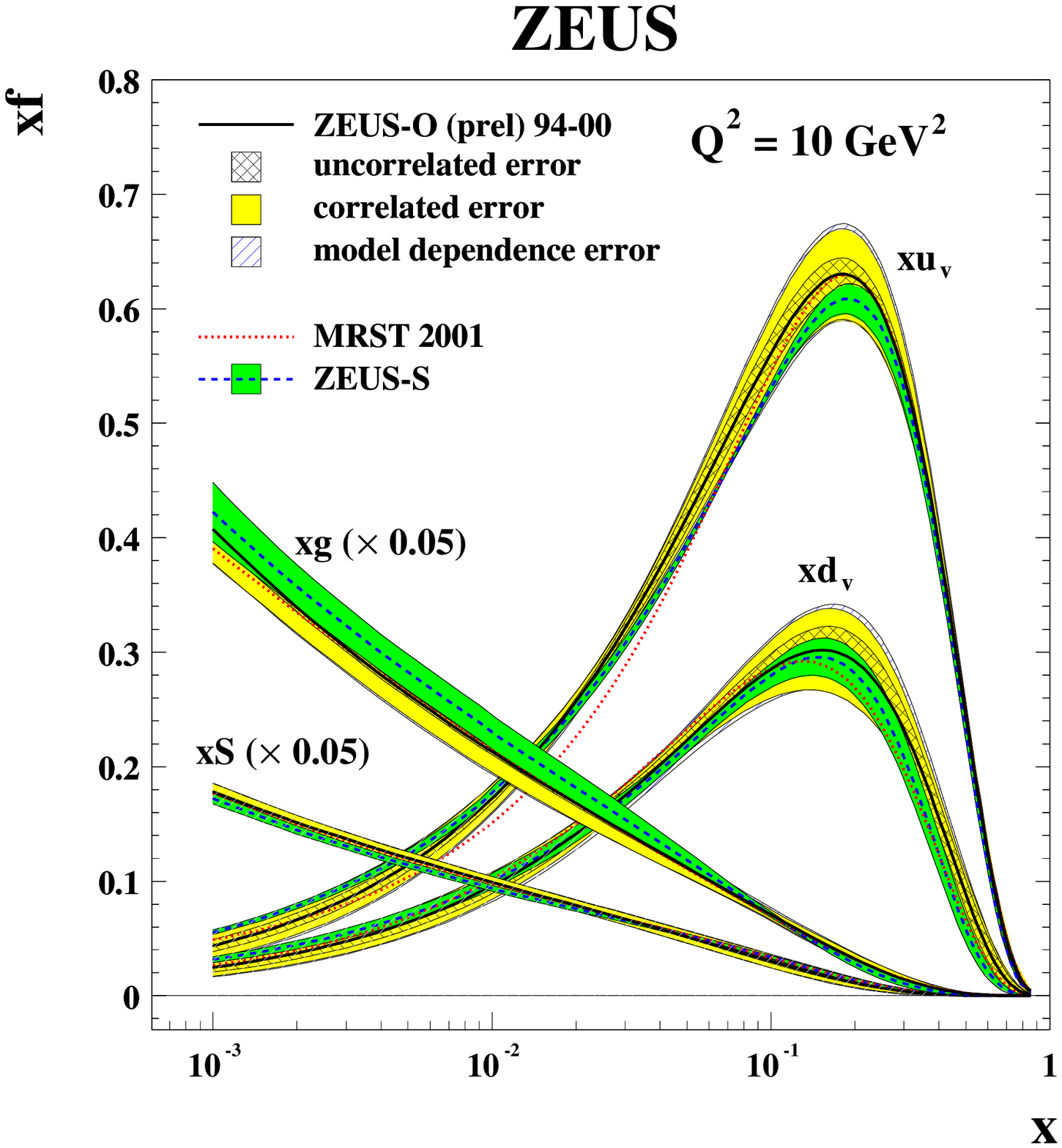,width={0.98\textwidth}}
\caption{New ZEUS-only PDFs fit using high-$Q^2$ data.}
\label{fig:QCDfit}
\end{minipage}
\end{center}
\end{figure}

ZEUS has measured the Charged and Neutral Current cross sections for
collisions of positrons and electrons on protons from the full HERA-I
data sample. Figure~\ref{fig:NCepem} shows the NC and CC cross sections as
a function of $Q^2$ for $e^+p$ and $e^-p$ scattering as measured by the
ZEUS~\cite{ZEUS} experiment. The ZEUS measurements are in a good
agreement with expectations of the Standard Model (SM), calculated with
the CTEQ6D~\cite{Kretzer:2003it} parameterization of parton distribution
functions (PDFs) in the proton.

Figure \ref{fig:QCDfit} shows results of the latest structure function
fit based on ZEUS data alone from the full HERA-I sample (ZEUS-O). The
addition of data used in the fit up to a $Q^2$ of 30,000 ${\rm GeV}^2$
reduces the need for fixed target data and sum rules to constrain the
fit for $x > 0.05$. There is good agreement between these new fits and
the previous ZEUS fits (ZEUS-S) from the 1994-1998 data sample, which
made more use of fixed target data, and with the MRST 2001
PDF\cite{MRST2001}.

\section{Hadronic Final States}

\begin{figure}
\begin{center}
\begin{minipage}[c]{0.48\textwidth}
\psfig{figure=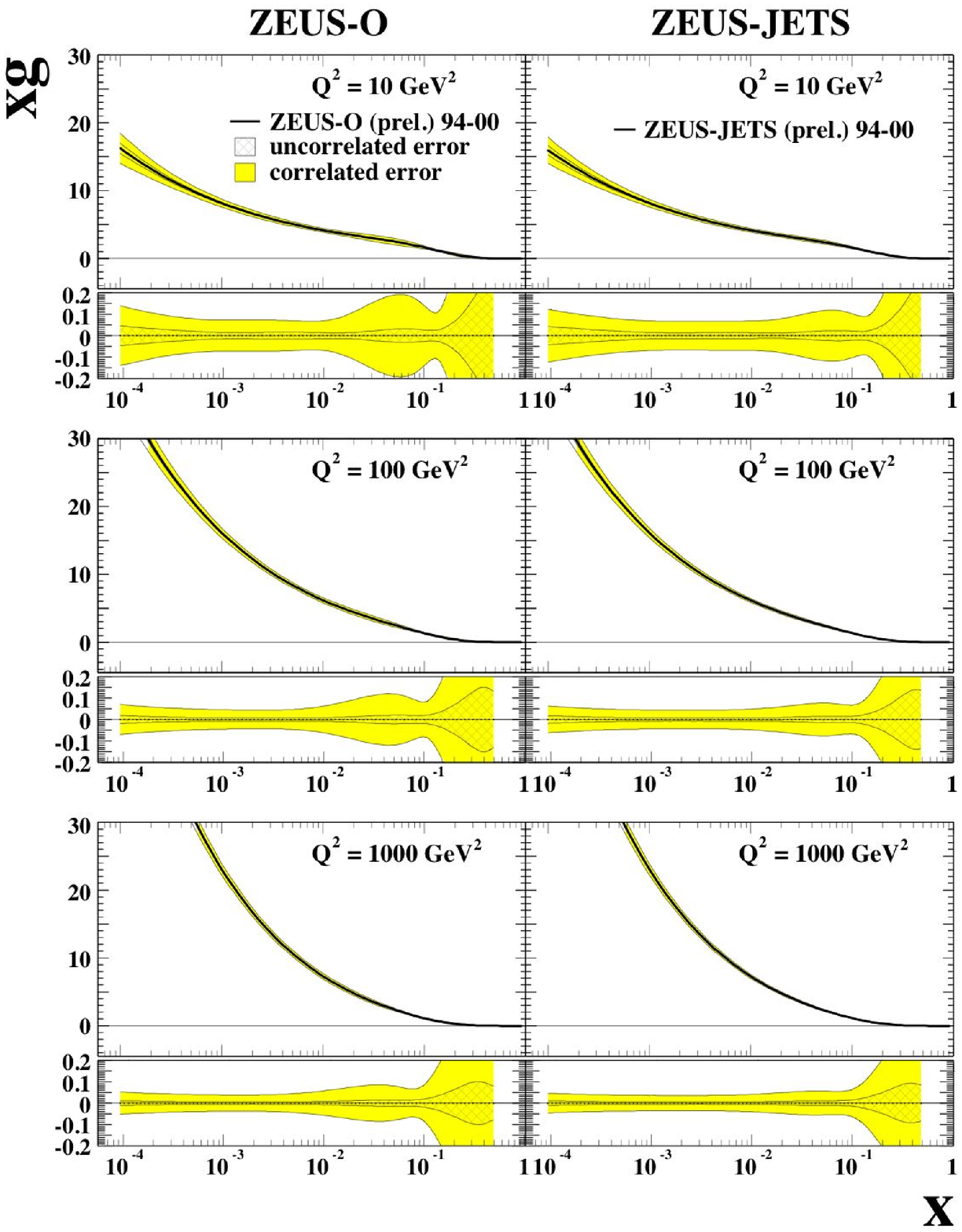, width={0.98\textwidth}}
\caption{ZEUS data only gluon QCD fits with (right) and
without (left) ZEUS inclusive  DIS and high $E_T$ dijet
photoproduction data incorporated.}
\label{fig:XGLU}
\end{minipage}
\hfill
\begin{minipage}[c]{0.48\textwidth}
\psfig{figure=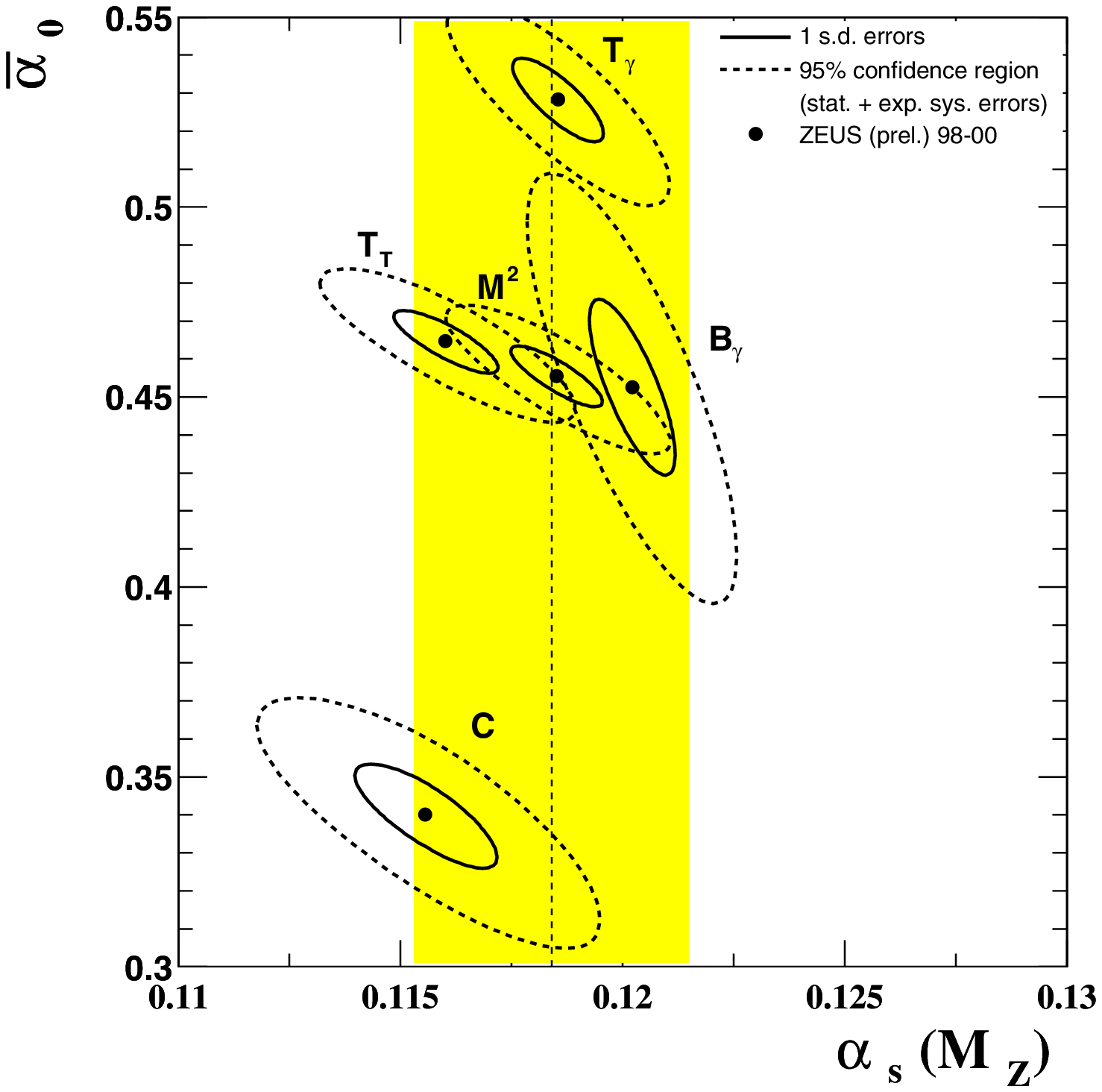, width={0.98\textwidth}}
\caption{$\alpha_s$ vs. $\bar \alpha_0 $ derived from
event shape variables measured in the current region
of the Breit frame.}
\label{fig:evtshape}
\end{minipage}
\end{center}
\end{figure}

The ZEUS collaboration has recently developed a rigorous
and consistent technique for including jets in QCD fits.  Figure \ref{fig:XGLU}
shows the effect of combining inclusive DIS jet data and high $E_T$ 
dijet photoproduction data into the ZEUS QCD fits. The figure shows the
ZEUS-only gluon fit with and without the jet data included in the fit. The
jet data constrains the medium-x gluon distribution, providing improved
precision on the gluon for $0.01 < x < 0.1$.

Event shape variables have been measured in the current region of the
Breit frame for NC DIS events. The Q-dependence of the means and
distributions of the shape variables has been compared with a model
based on fixed-order plus next-to-leading-logarithm perturbative
calculations and the Dokshitzer-Webber\cite{Dokshitzer:1995zt}
non-perturbative corrections (``power corrections''). The event shapes
examined and displayed in Fig. \ref{fig:evtshape} include thrust with
respect to the thrust ($T_T$) and $\gamma$ ($T_\gamma$) axes, broadening
($B_\gamma$), jet mass ($M^2$) and particle pair correlations (C). The
power corrections introduce the parameter $\bar \alpha_0$ to describe
the non-perturbative effects: $\left\langle F \right\rangle =
\left\langle F \right\rangle _{NLO} + \left\langle {F\left( {\bar \alpha
_0 } \right)} \right\rangle _{POW}$. The differential distributions are
fit with the combination of NLO QCD with power corrections and NLL
resummation\cite{Dasgupta:2001eq}, producing $\alpha _s \simeq 0.118,$
and $ \bar \alpha _0 \simeq 0.5$. The resummation extends the fit range
and yields consistent values of $\alpha_s$ and $\bar \alpha_0 $ for
$T_\gamma, T_T, B_\gamma,$ and $M^2$, with C displaying some
disagreement which depends on the fit range used.

\begin{figure}
\begin{center}
\begin{minipage}[c]{0.48\textwidth}
\psfig{figure=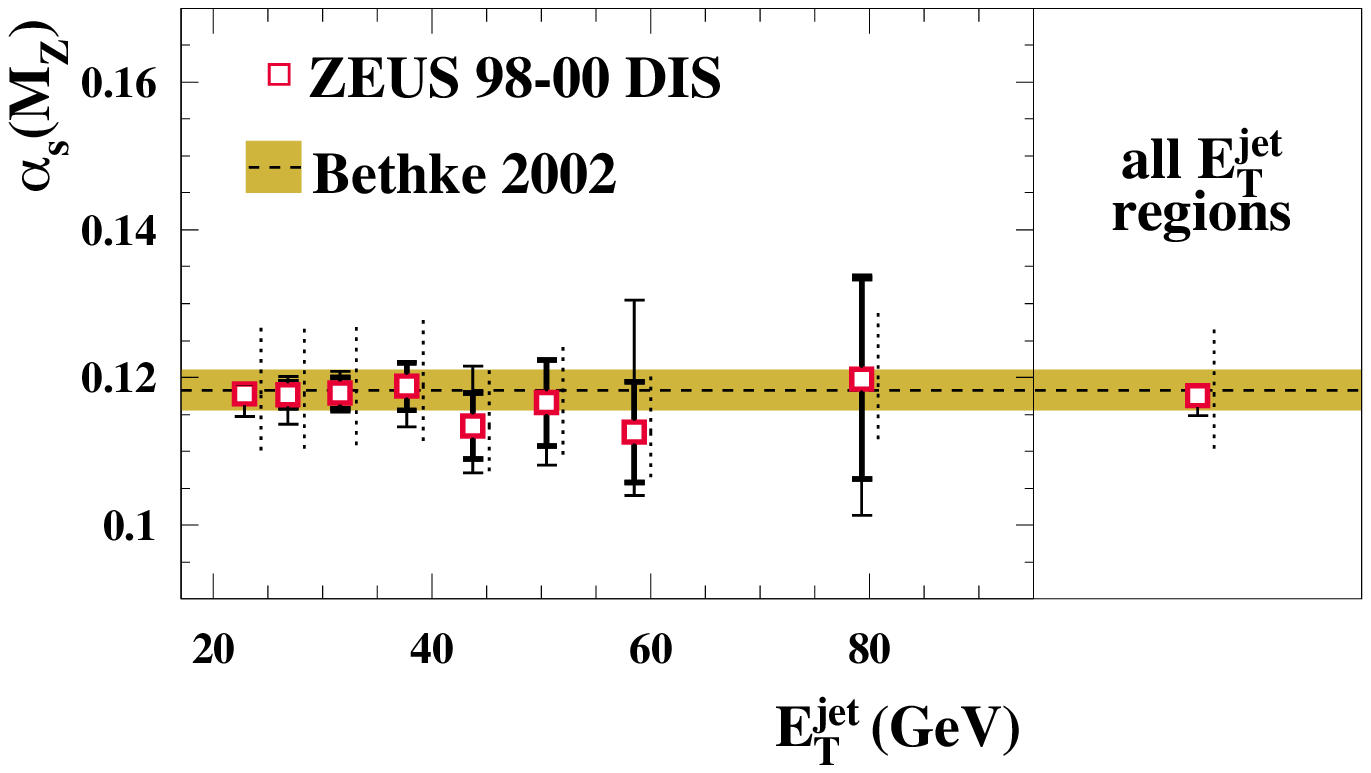, width={0.98\textwidth}}
\caption{The section of the plot on the right shows the $\as(\mz)$
values determined from the QCD fit of the measured integrated jet shape
$\langle \psi(r=0.5) \rangle$ in different $\etjet$ regions (squares).
The part on the right shows the combined value of $\as(\mz)$ obtained
using all the $\etjet$ regions (square). In both plots, the inner error
bars represent the statistical uncertainties of the data. The outer
error bars show the statistical and systematic uncertainties added in
quadrature. The dotted vertical bars represent the theoretical
uncertainties.}
\label{fig:jetshape}
\end{minipage}
\hfill
\begin{minipage}[c]{0.48\textwidth}
\psfig{figure=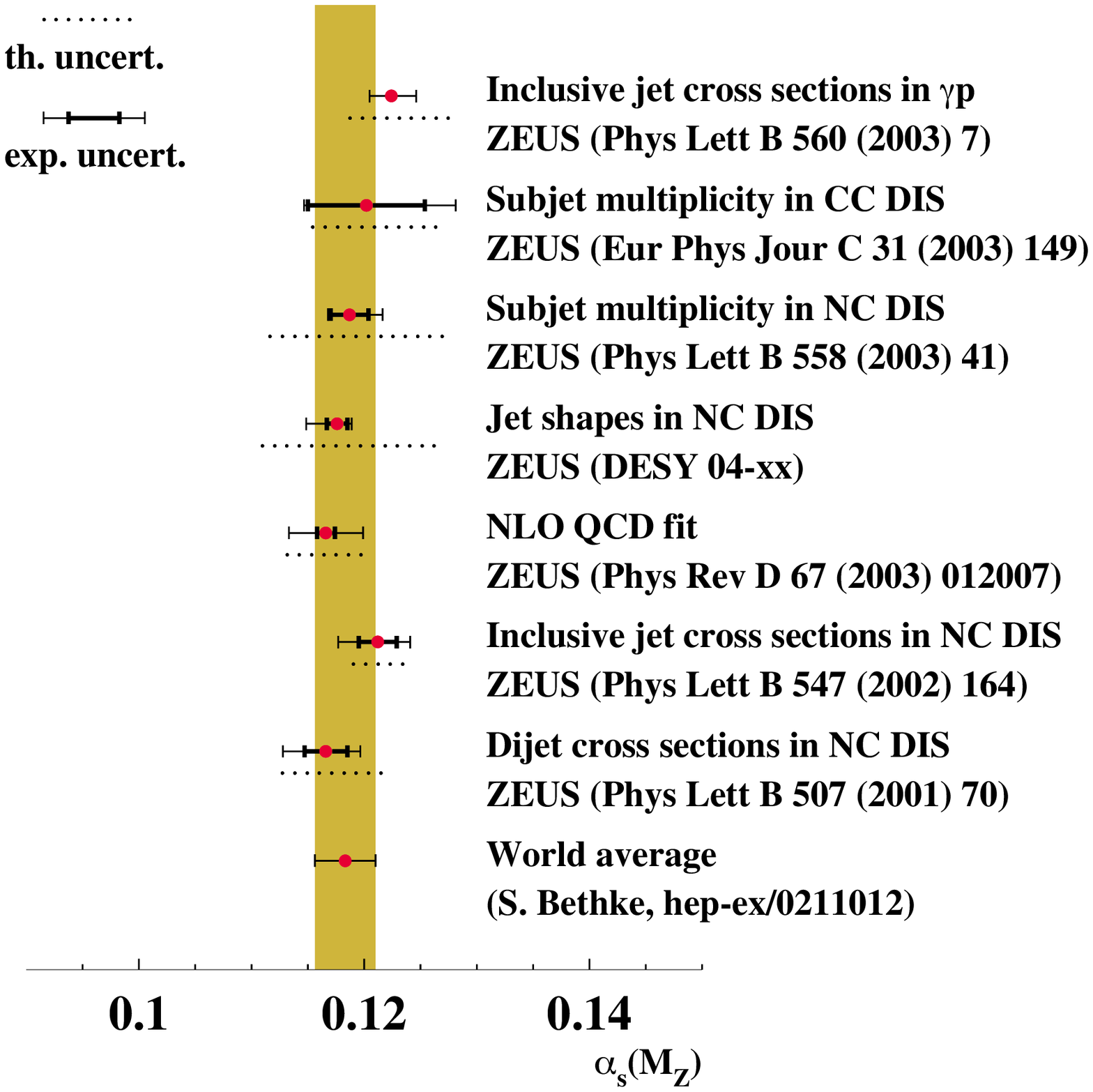, width={0.98\textwidth}}
\caption{$\as(\mz)$ values determined from ZEUS jet measurements,
and QCD fits to ZEUS structure functions compared with the world average.}
\label{fig:alphasum}
\end{minipage}
\end{center}
\end{figure}

The integrated jet shape, $\psi(r)$, is the fraction of the jet
transverse energy that lies inside a cone in the $\etaphi$ plane of
radius $r$ concentric with the jet axis while the mean integrated jet
shape, $\langle\psi(r)\rangle$, is defined as the averaged fraction of
the jet transverse energy inside the cone. Measurements of the mean
integrated jet shape in DIS have been used to extract a value of $\asz$
by comparing to the NLO QCD predictions of {\sc Disent}\cite{DISENT} as
a function of $\etjet$. The calculations reproduce the measured
observables well, demonstrating the validity of the description of the
internal structure of jets by pQCD. The values of $\asz$ as determined
from the measured $\langle\psi(r=0.5)\rangle$ in each region of $\etjet$
are shown in Fig.~\ref{fig:jetshape}. The value of $\asz$ as determined
by fitting the NLO QCD calculations to the measured mean integrated jet
shape $\langle\psi(r=0.5)\rangle$ for $\etjet>21$~GeV is
$\asmz{0.1176}{0.0009}{0.0026}{0.0009}{0.0072}{0.0091}$. This value is
in good agreement with the current world average. Comparisons of several
ZEUS measurements of $\asz$ with the current world average are shown in
Fig. \ref{fig:alphasum}. ZEUS is performing precision jet physics with
systematic errors at the level of 2\perc.

ZEUS has measured the differential dijet and trijet cross sections
in neutral  current  DIS for $ 10 < Q^2 < 5000~GeV^2 $ with high precision.
These cross sections have been measured as functions of the jet
transverse energy in the Breit Frame, $E^{jet}_{T,B}, \eta^{jet}_{LAB},$ and
$Q^2$ for jets with $E^{jet}_{T,B} > 5$ GeV and $-1 < \eta^{jet}_{LAB} <
2.5$. Events with two (three) or more jets were required to have the
invariant mass of the two (three) highest $E^{jet}_{T,B}$ jets to be
greater than 25 GeV to enable comparison with reliable NLO calculations.

\begin{figure}
\begin{center}
\begin{minipage}[c]{0.48\textwidth}
\psfig{figure=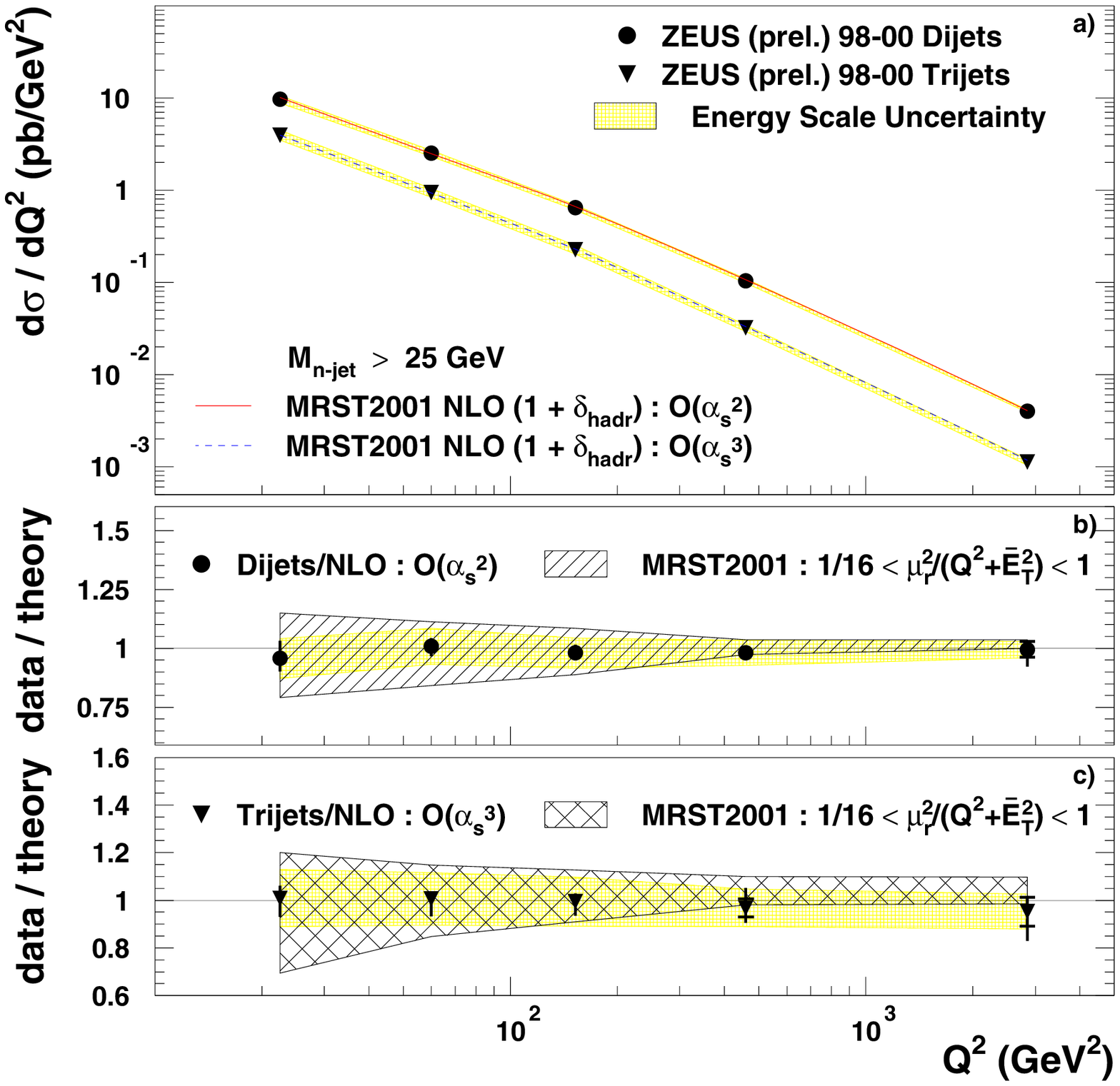, width={0.98\textwidth}}
\caption{a) The inclusive dijet and trijet cross sections as functions
of $Q^2$. the predictions of NLO pQCD is compared to the data. b) and c)
show the ratio of data over predictions. The hashed band represents the
renormalization scale uncertainty of the QCD calculation.}
\label{fig:di+trijet}
\end{minipage}
\hfill
\begin{minipage}[c]{0.48\textwidth}
\psfig{figure=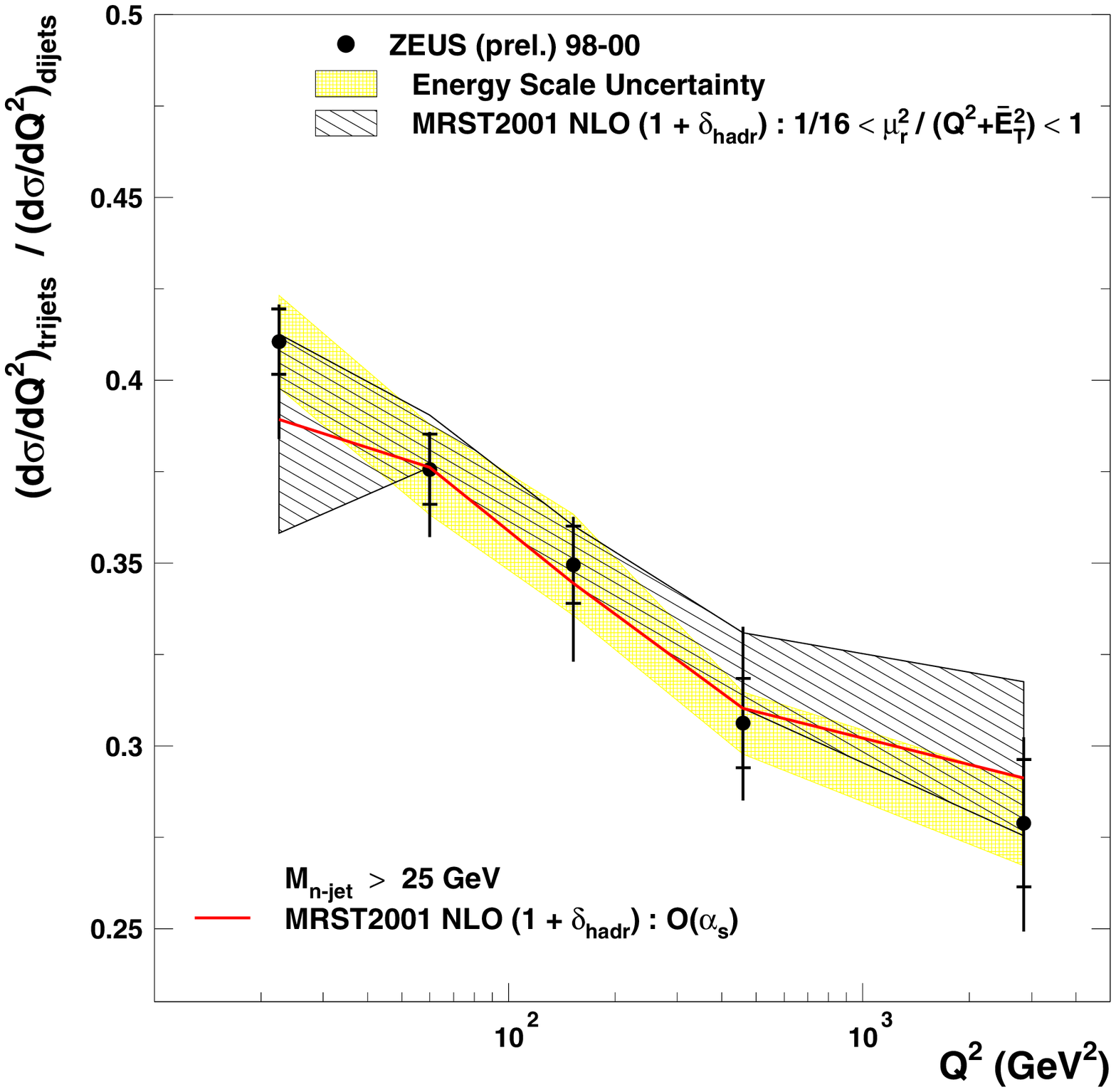, width={0.98\textwidth}}
\caption{The ratio of inclusive trijet over dijet cross section as a
function of $Q^2$. The predictions of NLO pQCD are compared to the data.
The hashed band represents the renormalization scale uncertainty of the
QCD calculation.}
\label{fig:tribydijet}
\end{minipage}
\end{center}
\end{figure}

The differential dijet and trijet cross sections as functions of $Q^2$
are presented in Fig. \ref{fig:di+trijet}. The largest uncertainty in
the data is due to absolute energy scale of the calorimeter. The cross
sections are compared with the NLOJET\cite{NLOJET} program using the
MRST2001\cite{MRST2001} PDFs and corrected for hadronization effects
using the {\sc Lepto}\cite{LEPTO} 6.5 program. Uncertainties due to terms
beyond NLO were estimated by varying both $\mu_R$ and $\mu_F$ between
$(\bar{E}^2_T + Q^2)$ and $(\bar{E}^2_T + Q^2)/16$. The data is well
described by the NLO QCD prediction. The renormalization scale error
grows larger at low $Q^2$.

Figure \ref{fig:tribydijet} shows the ratio of trijet to dijet cross
sections as a function of $Q^2$. The correlated systematic and
renormalization scale uncertainties cancel in this ratio. The agreement
between the data and the NLO predictions is good within the
uncertainties (experimental $\sim 5\perc$, theoretical $\sim 7\perc$),
which are substantially reduced from those of the dijet and trijet cross
sections alone and should be able to yield another precise measurement of
$\as(\mz)$. 

In the DGLAP formalism, the parton cascade that results from the hard
scattering of the virtual photon with a parton from the proton is
ordered in parton virtuality. This ordering along the parton ladder
implies an ordering in transverse energy of the partons, with the parton
participating in the hard scatter having the highest transverse energy.
In the BFKL formalism there is no strict ordering in virtuality and
transverse energy. Since the partons emitted at the bottom of the ladder
are closest in rapidity to the outgoing proton, they are manifested as
forward jets. BFKL evolution predicts that a larger fraction of small
$x$ events will contain high $E_T$ forward jets than is predicted by
DGLAP \cite{FORWARD}.

\begin{figure}
\begin{center}
\begin{minipage}[c]{0.48\textwidth}
\psfig{figure=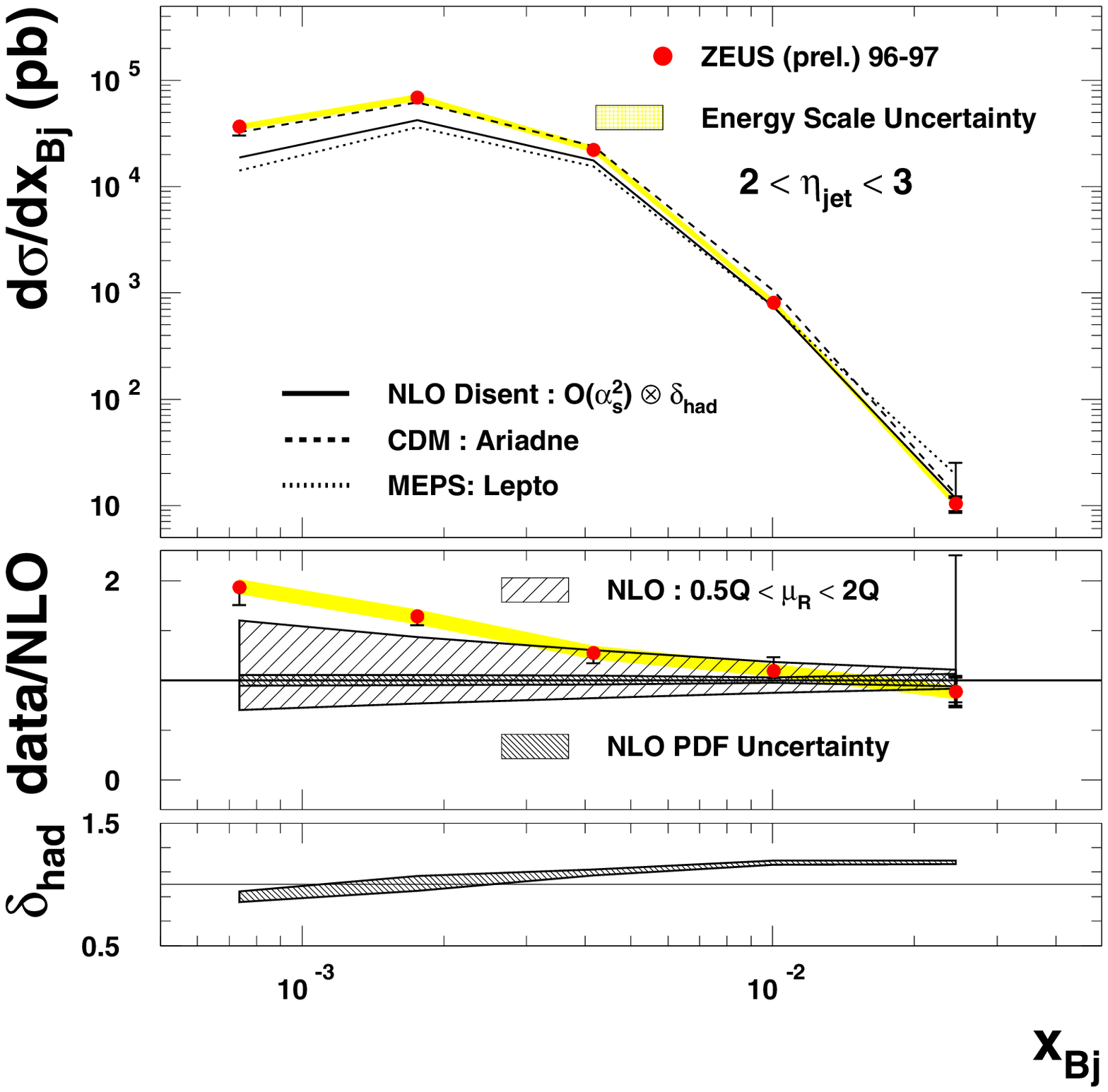, width={0.98\textwidth}}
\caption{Measured differential cross section (dots) in the forward BFKL
phase space for inclusive jet production in NC DIS with $ E _{T,jet} >
6$ GeV and $2 < \eta _{jet} < 3$ in the kinematic region defined by $Q^2
> 25~GeV^2$, $y > 0.04$ and $ \cos \gamma_h < 0$ as a function of
$x_{Bj}$. The lower section of each plot shows the hadronisation
correction factor applied to the QCD calculations.}
\label{fig:xBj_forw_eta}
\end{minipage}
\hfill
\begin{minipage}[c]{0.48\textwidth}
\psfig{figure=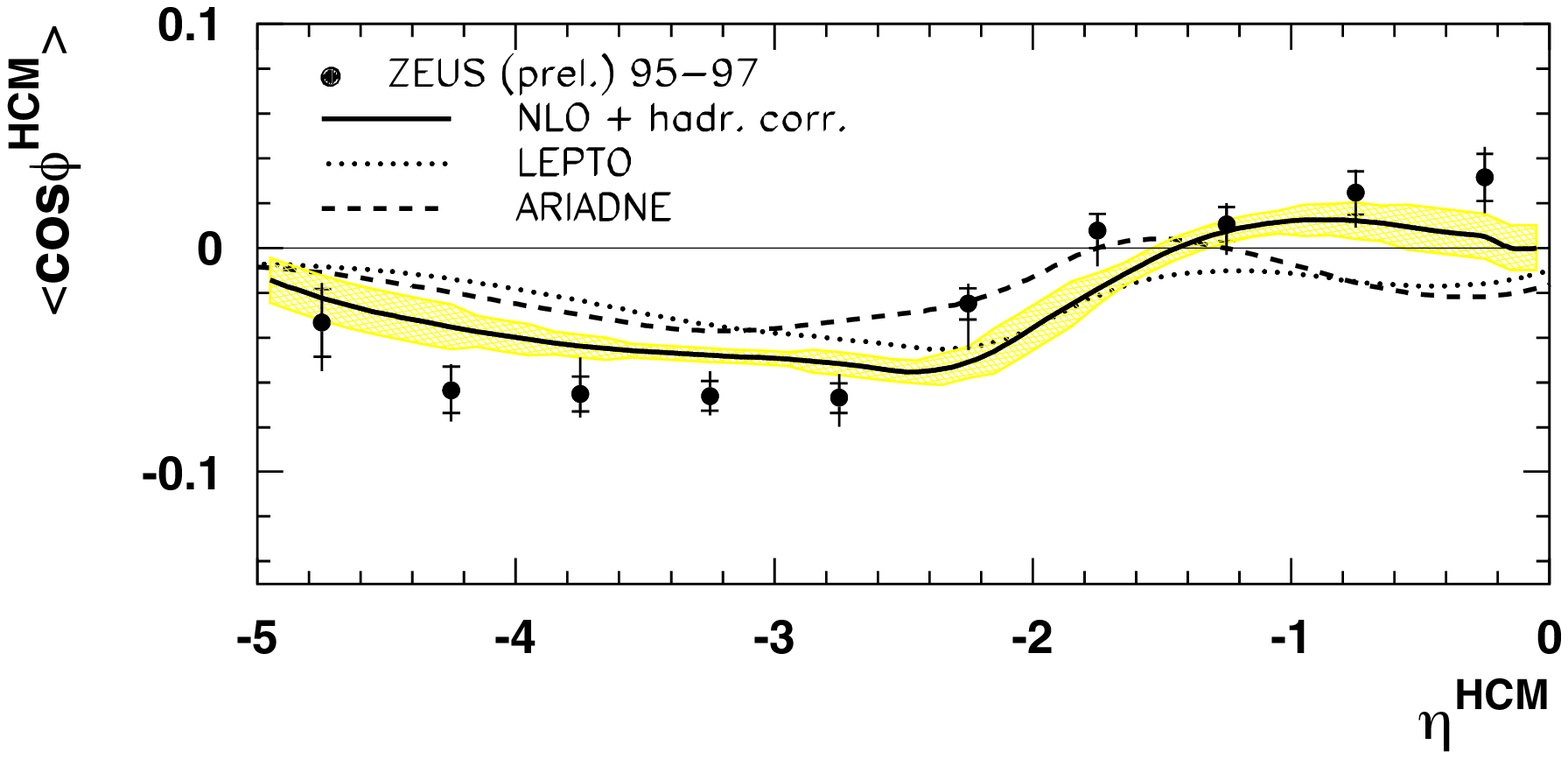, width={0.98\textwidth}}
\caption{ $\eta^{HCM}$ dependence of the azimuthal asymmetry of ZEUS
DIS data using energy flow objects in the hadronic center of mass frame
compared with LO MCs and a NLO calculation.}
\label{fig:asym_etacms}
\end{minipage}
\end{center}
\end{figure}

ZEUS has performed measurements of the DIS differential inclusive
forward jet cross sections in the lab frame with the specific
restrictions $ \cos \gamma _h < 0$ (where $\gamma _h$ is the angle that
defines the average direction of the hadronic system) to remove single
jets, $2 < \eta _{jet} < 3$ to select forward jets and $0.5 <
\frac{E_{T,jet}^2 }{Q^2 } < 2$ to limit the $Q^2$ evolution of the
particles on the parton ladder. This phase space region is where events
exhibiting BFKL effects are expected to be dominant. The measurements
are presented in Fig. \ref{fig:xBj_forw_eta}. The predictions of {\sc
Ariadne}\cite{ARIADNE} describe the data well, whereas the predictions
of {\sc Lepto}\cite{LEPTO} fail in all distributions. The {\sc Ariadne}
program uses the Color Dipole Model, which treats gluons emitted from
quark-antiquark (diquark) pairs as radiation from a color dipole
between two partons. This results in partons that are not ordered in
their transverse momenta.

The measurements are also compared with QCD predictions evaluated using 
the event generator {\sc Disent}\cite{DISENT}. 
The calculation describes the measurement as a function of
$x_{Bj}$ at high values, but underestimates the data in the low $x_{Bj}$ region
by nearly a factor of two. In this region there is a large renormalization uncertainty,
an indication of the importance of higher orders.

QCD predicts an asymmetry in the azimuthal distribution of hadrons
around the virtual photon direction in the hadronic center of mass frame
(HCM) as manifested in the distribution of the angle between the hadron
production plane and the electron scattering plane. QCD also predicts
that this azimuthal asymmetry will evolve in $\eta^{HCM}$ because the
contributions of the originating QCD-Compton and Boson Gluon fusion
processes vary with $\eta^{HCM}$. Figure \ref{fig:asym_etacms} shows the
ZEUS result for the $\eta^{HCM}$ dependence of this asymmetry from a new
analysis using energy flow objects in the hadronic center of mass frame.
This analysis includes charged and neutral hadrons where previous
analyses ony included charged tracks. It also enhances the contributions
from hard partons since these energy flow objects are weighted by
energy. The data are compared with LO MC ({\sc Lepto} and {\sc Ariadne})
and NLO calculations corrected to the hadron level by the LO MC. The NLO
effects are not negligible and provide better agreement with
experimental data.

\section{Heavy Flavors}

ZEUS has performed a search for pentaquarks in $\ksppb$  decay
channel \cite{zeuspq}. The analysis used deep inelastic scattering events
measured with exchanged-photon virtuality $Q^2\ge 1\gev^2$. The
data sample corresponded to an integrated luminosity of 121
pb$^{-1}$. The charged tracks were selected in the central
tracking (CTD) with $p_T\ge 0.15\gev$ and $|\eta|\le 1.75$,
restricting this study to a region where the CTD track acceptance
and resolution are high. 

\begin{figure}
\begin{center}
\begin{minipage}[c]{0.48\textwidth}
\psfig{figure=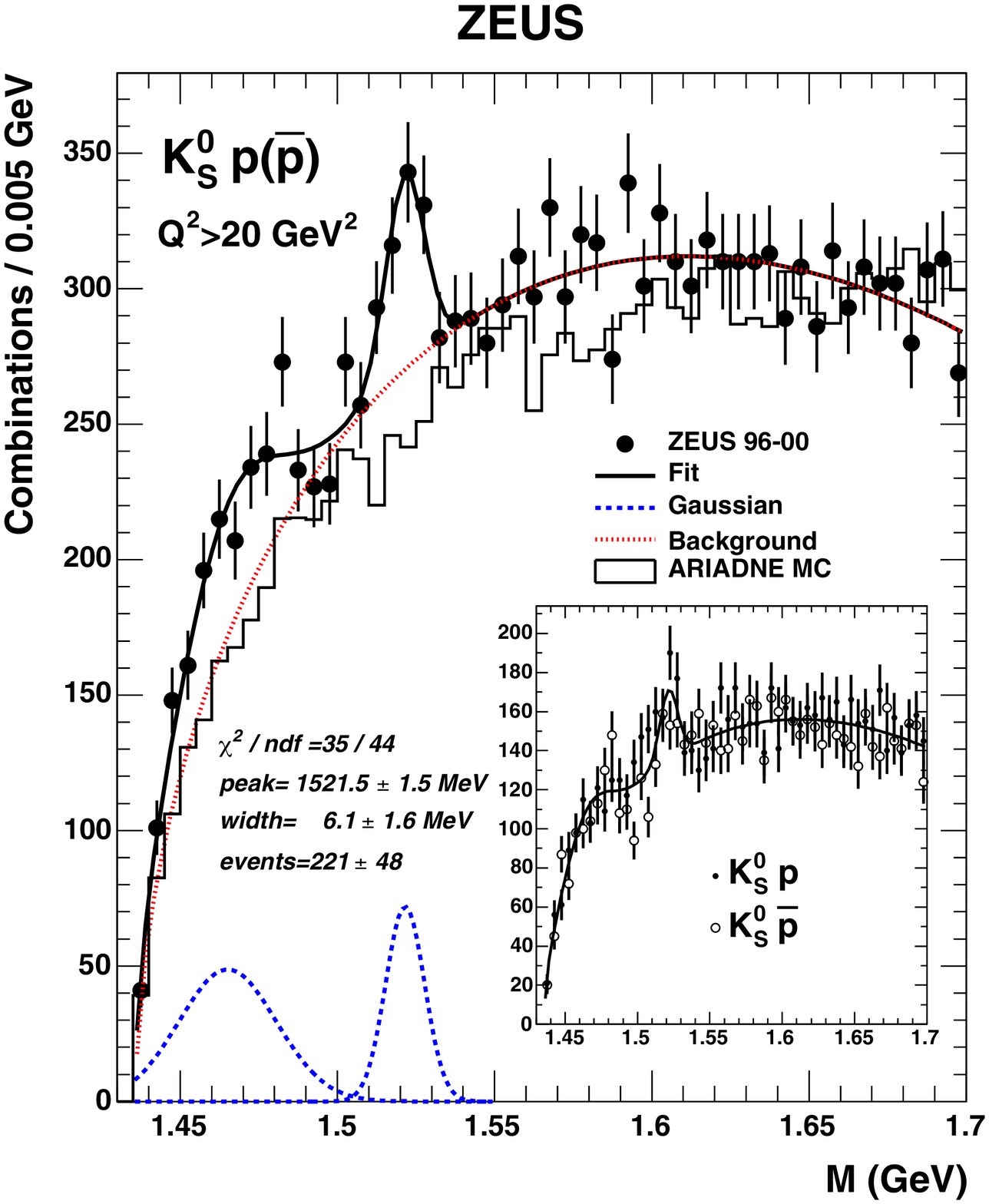, width={0.98\textwidth}}
\caption{Invariant-mass spectrum for the $\ksppb$ channel for
$Q^2 > 20 \gev^2$. The solid line is the result of a fit to the
data using the threshold background plus two Gaussians.  The
dashed lines show the Gaussian components, while the dotted line
indicates background. The prediction of the Monte Carlo
simulation is normalized to the data in the mass region above
$1650\mev$. The inset shows the $\kspb$ (open circles) and the
$\ksp$ (black dots) candidates separately, compared to the result
of the fit to the combined sample scaled by a factor of 0.5. }
\label{fig:zeuspq}
\end{minipage}
\hfill
\begin{minipage}[c]{0.48\textwidth}
\psfig{figure=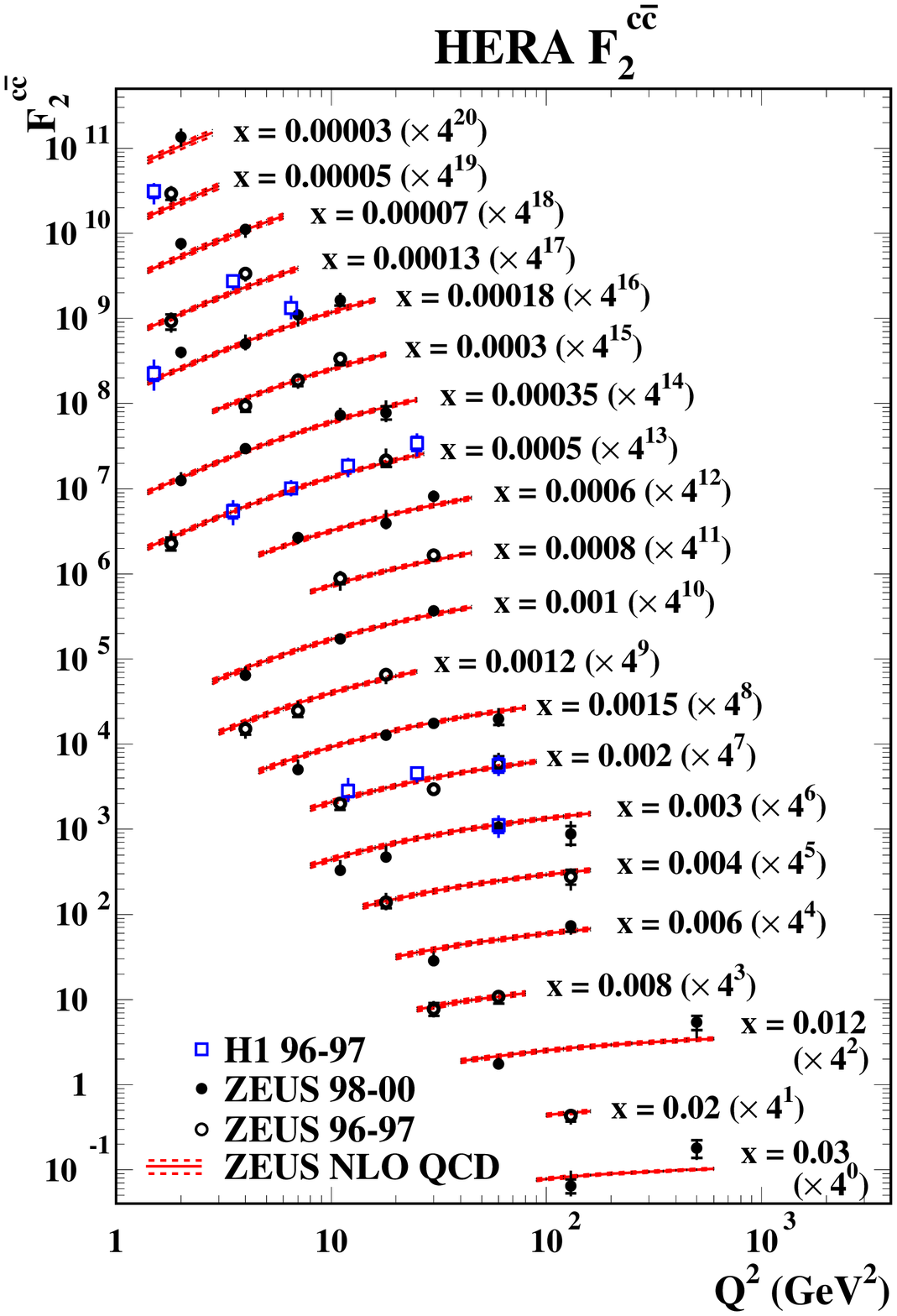, width={0.98\textwidth}}
\caption{The measured $F_2^{c \bar{c}}$ as a function of $Q^2$ for fixed
$x$ compared with previous H1 and ZEUS measurements and the
predictions from the ZEUS NLO QCD fit.}
\label{fig:f2cc_h1_zeus}
\end{minipage}
\end{center}
\end{figure}

Figure \ref{fig:zeuspq} shows the $\ksppb$ invariant mass for $Q^2>
20\gev^2$, as well as for the $\ksp$ and $\kspb$ samples separately
(shown as inset). The data is above the {\sc Ariadne} Monte Carlo model
near $1470\mev$ and $1522\mev$, with a clear peak at $1522\mev$. 

To extract the signal seen at $1522\mev$, the fit was performed using a
background function plus two Gaussians. The first Gaussian, which
significantly improves the fit at low masses, may correspond to the
unestablished PDG $\Sigma (1480)$. The peak position determined from the
second Gaussian was $1521.5\pm 1.5({\rm stat.})^{+2.8}_{-1.7} ({\rm
syst.}) \mev$. It agrees well with the measurements by HERMES, SVD and
COSY-TOF for the same decay channel \cite{ks}. If the width of the
Gaussian is fixed to the experimental resolution, the extracted
Breit-Wigner width of the signal is $\Gamma=8\pm 4(\mathrm{stat.})
\mev$.

The number of events ascribed to the signal by this fit was $221 \pm
48$. The statistical significance, estimated from the number of events
assigned to the signal by the fit, was $4.6\sigma$. The number of events
in the $\kspb$ channel was $96 \pm 34$. It agrees well with the signal
extracted for the $\ksp$ decay mode. If the observed signal corresponds
to the pentaquark, this provides the first evidence for its antiparticle
with a quark content of $\bar{u}\bar{u}\bar{d}\bar{d}s$.

ZEUS has measured the open-charm contribution, $F_2^{c \bar{c}}$, to
the proton structure-function $F_2$. This final HERA-I result is based
on 5500 $D^*$, providing 31 additional data points over the previous
ZEUS result~\cite{epj:c12:35} and increasing the range up to $Q^2 =
500~{\rm GeV}^2$. The data for $F_2^{c \bar{c}}$ as a function of $x$ for
fixed $Q^2$ are shown in Figure~\ref{fig:f2cc_h1_zeus} to be consistent
with previous measurements from H1~\cite{pl:b528:199} and
ZEUS~\cite{epj:c12:35} and with the recent ZEUS NLO
fit~\cite{pr:d67:012007}. The data rise with increasing $Q^2$, with the
rise becoming steeper at lower $x$, demonstrating the property of
scaling violation in charm production. The uncertainty on the
theoretical prediction is that from the PDF fit propagated from the
experimental uncertainties of the fit data. At the lowest $Q^2$, the
uncertainty in the data is comparable to the PDF uncertainty shown. This
implies that the double-differential cross sections could be used as an
additional constraint on the gluon density in the proton.

\begin{figure}
\begin{center}
\begin{minipage}[c]{0.48\textwidth}
\psfig{figure=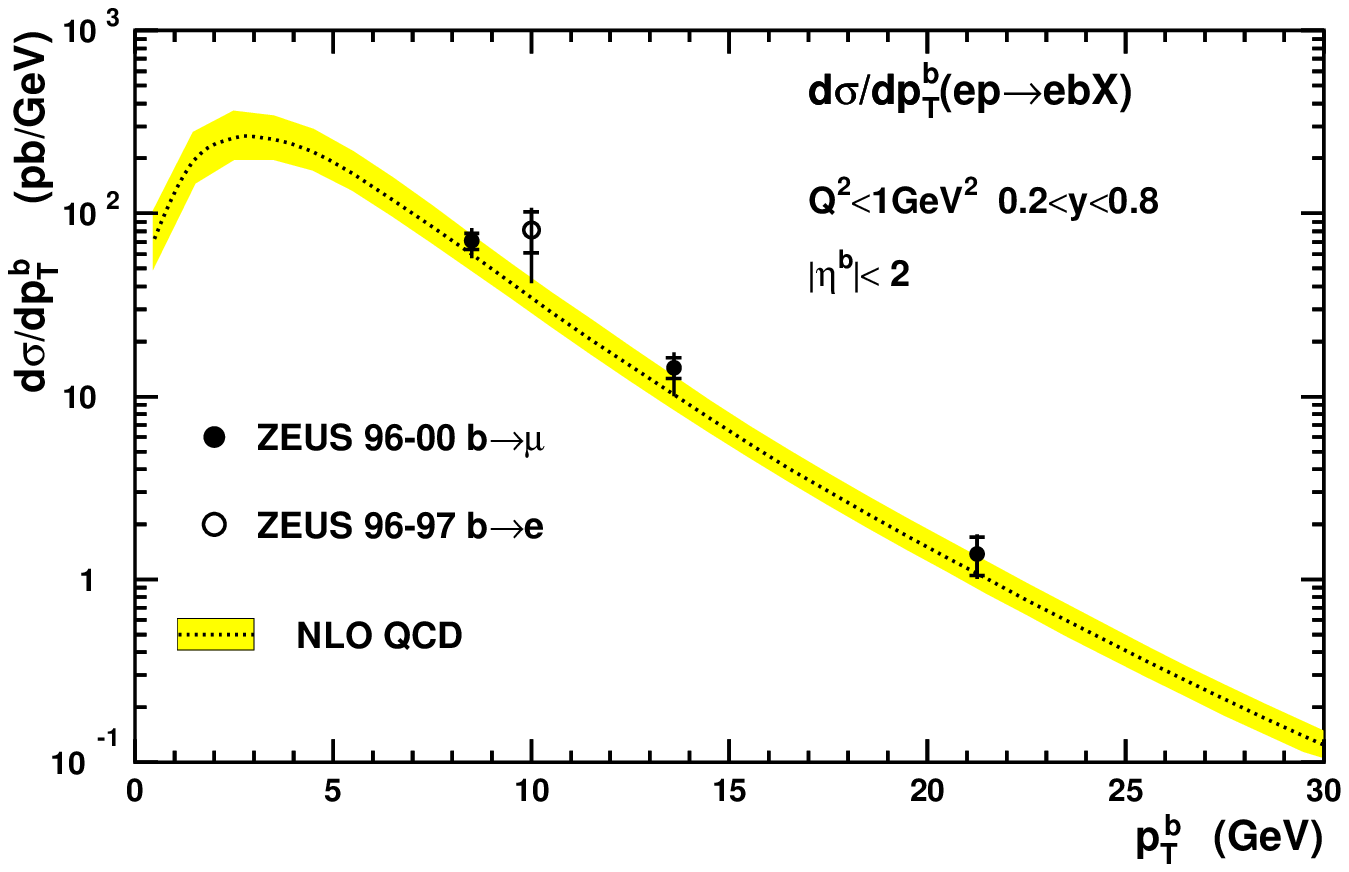, width={0.98\textwidth}}
\caption{Differential cross section for $b$-quark production as a
function of the $b$-quark transverse momentum $p_{T}^b$ for $b$-quark
pseudorapidity $|\eta^b|<2$ and for $Q^2<1 \gev^2$, $0.2<y<0.8$. The
filled points show the new ZEUS results and the open point is the
previous ZEUS measurement in the electron channel \cite{ZEUSelec}. The
full error bars are the quadratic sum of the statistical (inner part)
and systematic uncertainties. The dashed line shows the NLO QCD
prediction with the theoretical uncertainty shown as the shaded band.}
\label{fig:beautyphpcross}
\end{minipage}
\hfill
\begin{minipage}[c]{0.48\textwidth}
\psfig{figure=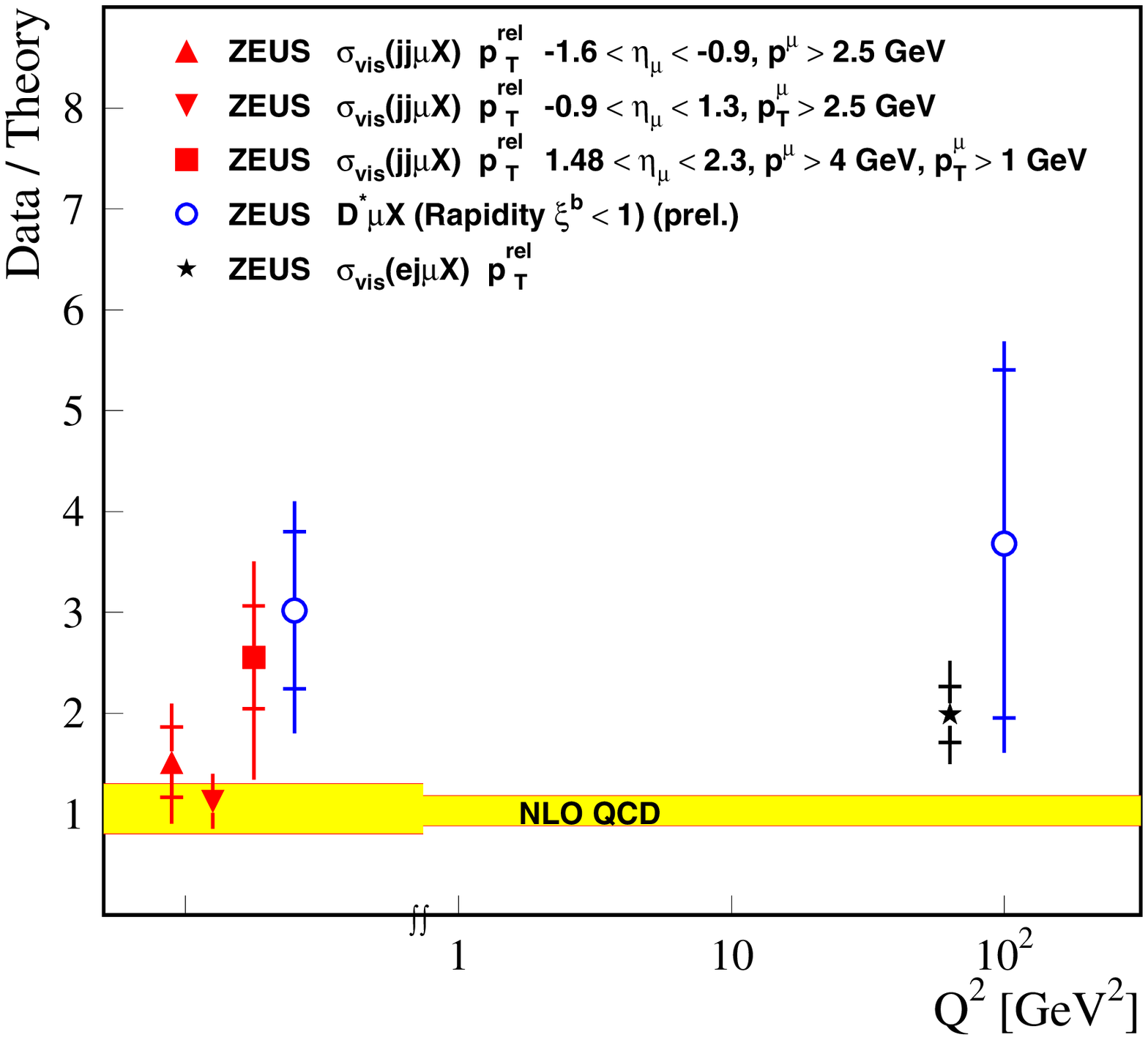, width={0.98\textwidth}}
\caption{The ratio of ZEUS data to a NLO QCD calculation as a function
of $Q^2$ for differential b-quark cross sections measured in
photoproduction and DIS events with at least one jet reconstructed in
the Breit frame and a muon. The NLO shaded band shows the uncertainty
due to renormalization and factorization scales and b-quark mass.}
\label{fig:beauty_sum}
\end{minipage}
\end{center}
\end{figure}

ZEUS has measured beauty photoproduction in events with two jets and a
muon, for $Q^2<1 \gev^2$ \cite{beautyphp}. These events have been shown
to agree with NLO QCD predictions based on the FMNR \cite{fmnr} program.
This NLO QCD prediction was used to extrapolate the cross section for
dijet events with a muon to the inclusive $b$-quark cross section. The
$b$-quark differential cross section was measured as a function of the
quark transverse momentum, $d\sigma(ep \rightarrow b X)/dp_{T}^b$, for
$b$-quark pseudorapidity in the laboratory frame $|\eta^b|<2$. The
result, shown in Figure \ref{fig:beautyphpcross}, is consistent with the
previous ZEUS result from semi-leptonic B decays into electrons
\cite{ZEUSelec} translated into the $b$-quark cross section for
$p_{T}^b> p_{T}^{\rm min} = 5 \gev$ and $|\eta^b|<2$, converted to a
differential cross section using the NLO prediction and plotted at the
average $b$-quark transverse momentum. Beauty photoproduction in $ep$
collisions is reasonably well described by NLO QCD.

ZEUS has also measured the b cross section in DIS events with at least
one hard jet in the Breit frame ($E_T^{Breit} > 6 GeV$) together with a
muon and electron for photon virtualities $Q^2 > 2 {\rm GeV}^2$, as well
as from events with a muon and a D*. These measurements have been
compared with an NLO QCD calculation in the HVQDIS program\cite{HVQDIS}.
Figure \ref{fig:beauty_sum} compares the cross sections from these
measurements and the photoproduction measurements in various kinematic
regions shown with the NLO QCD calculation. While, as mentioned above,
the photoproduction measurements are in good agreement with the NLO QCD
calculation, the DIS measurements, although in general consistent with
the NLO QCD calculation, are about two standard deviations above the NLO
QCD calculation at low values of $Q^2$, Bjorken $x$ and muon transverse
momentum, and high values of jet transverse energy and muon
pseudorapidity.

\section{Diffraction and Vector Meson Production}

\begin{figure}
\begin{center}
\begin{minipage}[c]{0.48\textwidth}
\psfig{figure=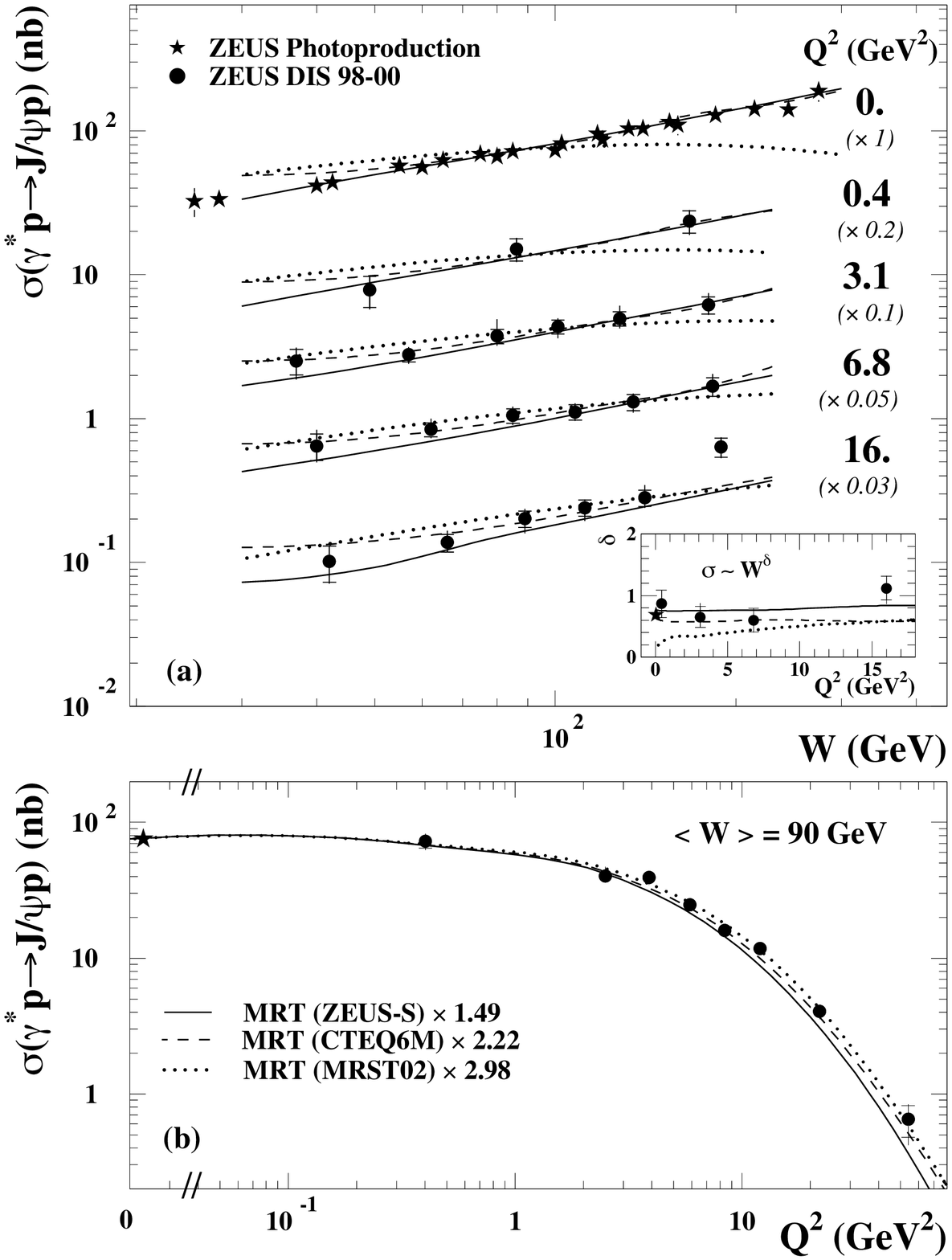, width={0.98\textwidth}}
\caption{Exclusive $J/\psi$ electroproduction cross section (a) as a
function of $W$ for four values of $Q^2$ and (b) as a function of $Q^2$
at $\langle W\rangle=90\gev$. ZEUS photoproduction results are also
shown. The data are compared to the MRT predictions (see text) obtained
with different parametrisations of the gluon density and normalized to
the ZEUS photoproduction point at $\langle W\rangle=90\gev$. The insert
shows the parameter $\delta$ as a function of $Q^2$. The inner error
bars represent the statistical uncertainties, the outer bars are the
statistical and systematic uncertainties added in quadrature. An overall
normalisation uncertainty of $^{+5\%}_{-8\%}$ was not included.}
\label{fig:jpsidiff}
\end{minipage}
\hfill
\begin{minipage}[c]{0.48\textwidth}
\psfig{figure=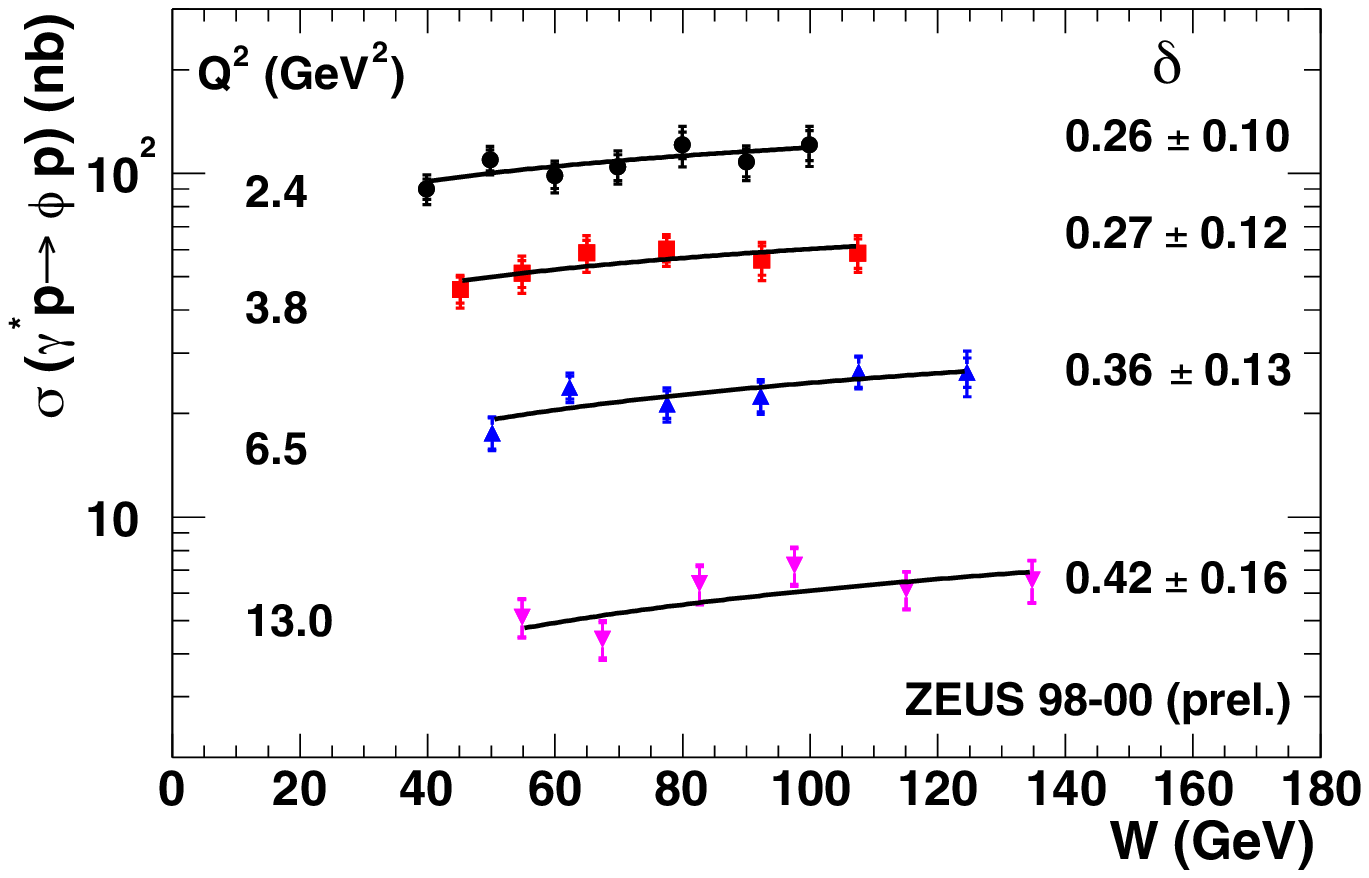, width={0.98\textwidth}}
\psfig{figure=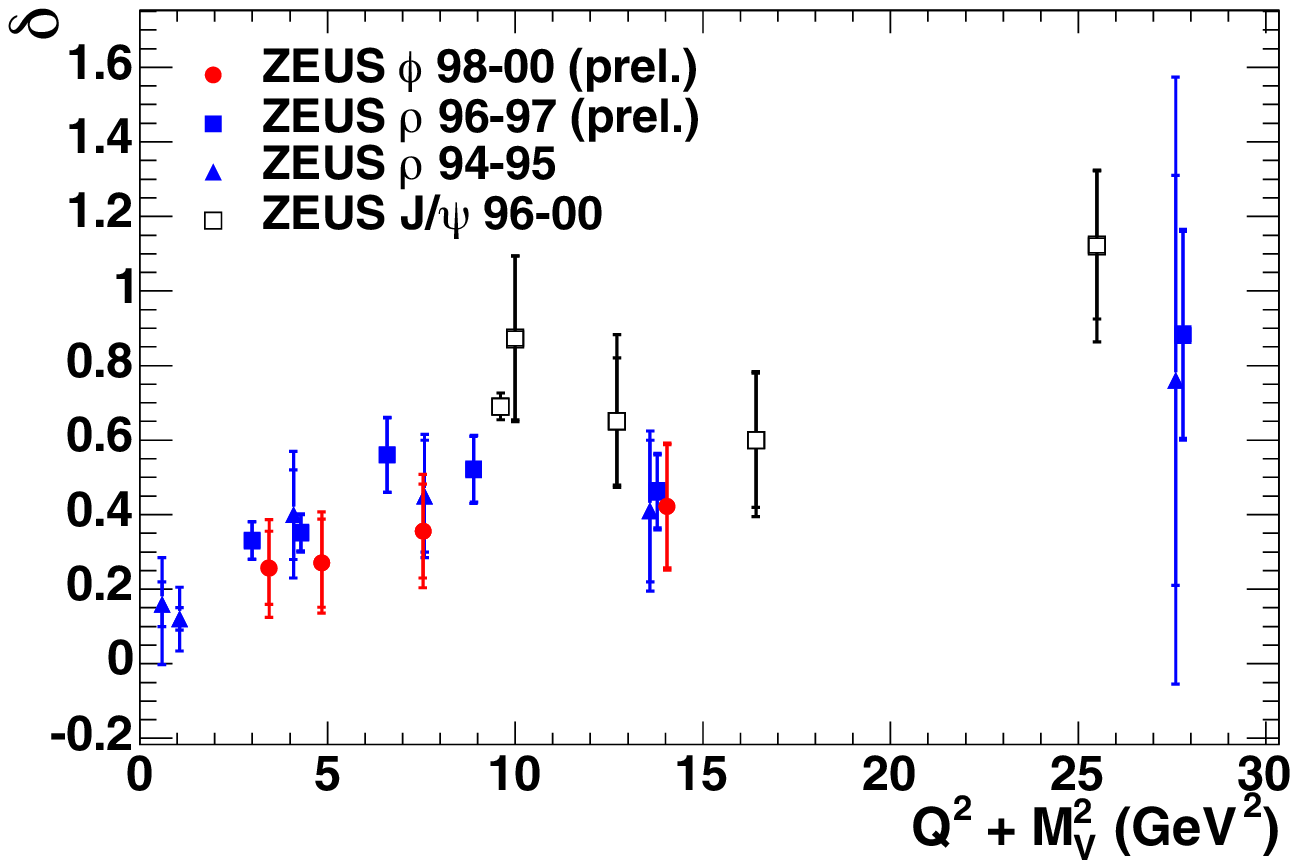, width={0.98\textwidth}}
\caption{(top) Exclusive $\phi$ cross section as a function of $W$ for
four values of $Q^2$. The solid lines are a result of a fit to the form
$\sigma \propto W^{-\delta}$. (bottom) Extracted values of $\delta$
compared with results from other VMs. The inner error bars represent
the quadratic sum of the statistical uncertainties, while the outer bars
represent the quadratic sum of the statistical and systematic
uncertainties. The overall normalization uncertainty of
$^{+10.4}_{-6.6}\perc$ is not included in the error bar in the upper
plot.}
\label{fig:phivwq}
\end{minipage}
\end{center}
\end{figure}

ZEUS has measured the exclusive electroproduction of $J/\psi$ mesons,
$ep\rightarrow e J/\psi\:p$, for photon virtualities in the ranges
$0.15<Q^2<0.8\gev^2$ and $2<Q^2<100\gev^2$, for photon-proton
center-of-mass energies in the range $30<W<220\gev$. The cross section
of the process $\gamma^*\:p\rightarrow J/\psi\:p$ measured for
$|t|<1\gev^2$, but extrapolated to the full $t$ range, rises with $W$ as
$\sigma \propto W^{\delta}$, with a slope parameter $\delta$ of about
0.7. This parameter does not change significantly with $Q^2$ and is
consistent with that observed in $J/\psi$ photoproduction. The cross
section at $W=90\gev$ and over the whole $Q^2$ range is described by the
function $\sigma\propto (Q^2+M^2_{J/\psi})^{-n}$, with $n=2.44\pm0.08$,
and is plotted in Fig. \ref{fig:jpsidiff} as a function of $W$ and
$Q^2$, along with the ZEUS measurement of exclusive $J/\psi$
photoproduction~\cite{epj:c24:345}. The data is also compared to the LO
calculation of Martin, Ryskin and Teubner (MRT)~\cite{pr:d62:14022} with
three different gluon distributions: MRST02~\cite{epj:c23:73},
CTEQ6M~\cite{JHEP:0207:012} and ZEUS-S~\cite{pr:d67:012007}, obtained
from NLO DGLAP analyses of structure function data and normalized to the
ZEUS photoproduction measurement at $W=90\gev$. While CTEQ6M describes
the $W$ and $Q^2$ dependence of the data, MRST02 has the wrong shape in
$W$, particularly at low $Q^2$. ZEUS-S describes the $W$ dependence but
falls too quickly with increasing $Q^2$. The data exhibit a strong
sensitivity to the gluon distribution in the proton. However, full NLO
calculations are needed in order to use these data in global fits to
constrain the gluon density. 

ZEUS has measured the exclusive electroproduction of $\phi$ mesons,
$ep\rightarrow e \phi \:p$, for photon virtualities in the range and
$2<Q^2<70\gev^2$, for photon-proton center-of-mass energies in the range
$35<W<145\gev$ and $|t|<0.6\gev^2$. Figure \ref{fig:phivwq} (top) shows
the extracted cross section of the process $\gamma^*\:p\rightarrow \phi
\:p$ measured as a function of $W$ and $Q^2$. The data were fit to a
dependence $\sigma\propto W^\delta$ with $\delta \approx$ 0.3. The fit
values of $\delta$ are compared to those from previous ZEUS measurements
for different VM in Fig. \ref{fig:phivwq} (bottom). The behavior of
$\delta$ is consistent with a scaling of the W dependence for exclusive
VM production with $Q^2 + M^2_V$.

\begin{figure}
\begin{center}
\begin{minipage}[c]{0.48\textwidth}
\psfig{figure=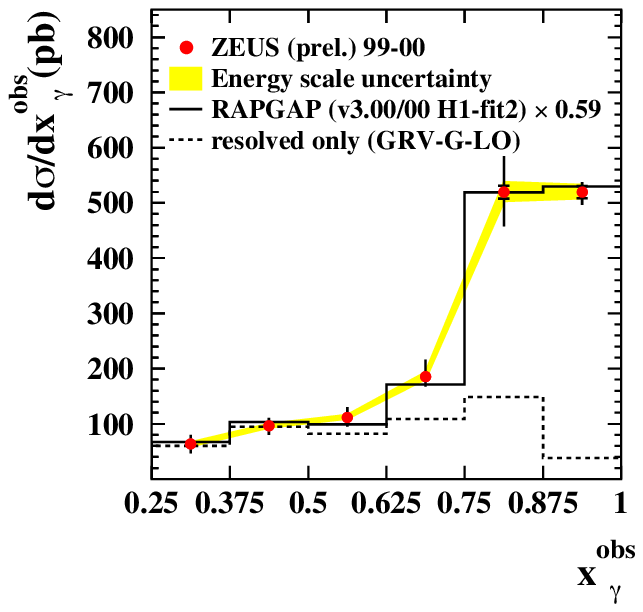, width={0.98\textwidth}}
\caption{The single differential cross section in $x^{obs}_\gamma$. The
data is shown as dots with the corresponding energy scale uncertainty shown
as a band; the inner error bards indicate the statistical uncertainty and the
outer error bars indicate the statistical and systematic errors added in quadrature.
The solid lines show the prediction of the LO {\sc Rapgap} MC, normalized to the
data by a factor of 0.59; the dashed line is the resolved photon component
from {\sc Rapgap}}. 
\label{fig:diffdijet}
\end{minipage}
\hfill
\begin{minipage}[c]{0.48\textwidth}
\psfig{figure=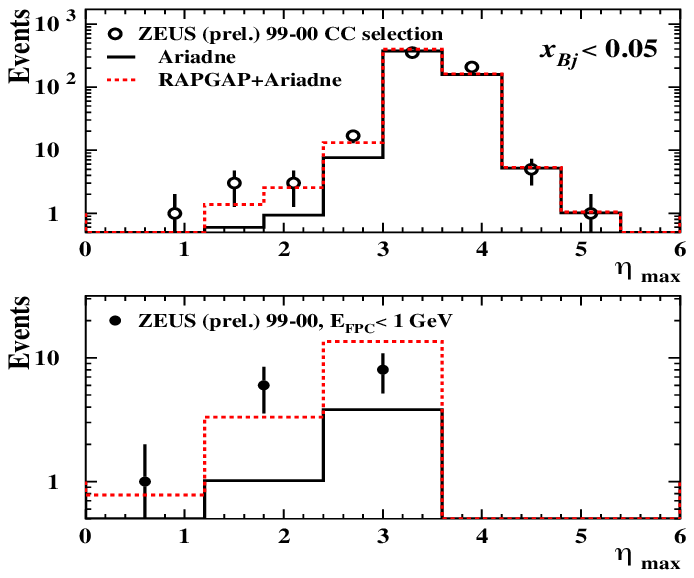, width={0.98\textwidth}}
\caption{(top) $\eta_{max}$  distribution for ZEUS CC DIS data (dots), the
non-diffractive {\sc Ariadne} and {\sc Ariadne} plus {\sc Rapgap}
and (bottom) the same plot with the additional $E_{FPC} < 1$ cut.}
\label{fig:ccgap}
\end{minipage}
\end{center}
\end{figure}

ZEUS has studied whether hard diffractive processes can be factorized
into universal diffractive PDFs and the partonic cross section using a
data sample from 30 times the luminosity of the previous ZEUS
result\cite{EPJC5} and using a new forward detector added near the beam
pipe, that allows measurements to be made over a wider kinematic range.
Dijet events are selected with $ 0.2 < y < 0.85, Q^2 < 1.0 {\rm GeV}^2, x_{pom} <
0.035, E_T^{jet1(2)} > 7.5 (6.5)$ GeV and $ -1.5 < \eta^{jet1,2} < 2.0$.
Diffractive events are selected with a rapidity gap in the forward
region, $ 3 < \eta < 5$. Figure \ref{fig:diffdijet} shows that the LO
prediction from the {\sc Rapgap}\cite{RAPGAP} generator, normalized to
the data by a factor of 0.59 describes the $x_\gamma^{obs}$ well in both
direct (high $x_\gamma^{obs}$) and resolved (low $x_\gamma^{obs}$)
enriched regions. A reduction in the data below the prediction at lower
$x_\gamma^{obs}$ would suggest a suppression of the resolved photon
processes with respect to the direct photon processes, indicating a
breakdown of QCD factorization, but this is not observed.

ZEUS has studied CC DIS events with a Large Rapidity Gap (LRG) for $Q^2
> 200 {\rm GeV}^2$ and $x_{Bj} > 0.05$. Figure \ref{fig:ccgap} shows the
distribution of $\eta_{max}$, where $\eta_{max}$ is the pseudorapidity
of the energy distribution in the calorimeter above 400 MeV which is
closest to the outgoing proton direction. The data are compared with a
combination of non-diffractive events produced by {\sc
Djangoh}\cite{DJANGOH} interfaced to {\sc Ariadne}\cite{ARIADNE} for
fragmentation and diffractive events modeled by the {\sc
Rapgap}\cite{RAPGAP} generator. The upper plot in Fig. \ref{fig:ccgap}
shows an excess of events with a LRG over {\sc Ariadne} at low
$\eta_{max}$. The bottom plot in Fig. \ref{fig:ccgap} shows the
$\eta_{max}$ distribution with the requirements that the Forward Plug
Calorimeter (extending the acceptance of the forward calorimeter by one
unit to $\eta = 5$) satisfies $E_{FPC} < 1$ GeV, suppressing the
non-diffractive contribution. Additional cuts of $\eta_{max} < 2.9$ and
$x_{pom} < 0.05$ (where $x_{pom}$ is the fraction of proton longitudinal
momentum carried by the diffractive exchange) further reduce the
non-diffractive background, resulting in 9 data events, $5.6 \pm 0.7$
diffractive CC ({\sc Rapgap}) simulated events and $2.1 \pm 0.4$
non-diffractive CC ({\sc Ariadne}) simulated events.

The assumption that the excess of LRG
events over the non-diffractive expectation from {\sc Ariadne} is due to
diffraction results in the cross section, $\sigma _{(Q^2 > 200GeV^2
,x_{pom} < 0.05)}^{CC\;DIFF}$ = $0.49 \pm 0.20 (stat.) \pm 0.13 (syst.)$
pb, compared with a prediction of 0.4 pb from {\sc Rapgap}. The
ratio of diffractive to total CC cross sections,
\[
\begin{array}{l}
 {{\sigma _{(Q^2  > 200GeV^2 ,x_{pom}  < 0.05)}^{CC\;DIFF} } \mathord{\left/
 {\vphantom {{\sigma _{(Q^2  > 200GeV^2 ,x_{pom}  < 0.05)}^{CC\;DIFF} } {\sigma _{(Q^2  > 200GeV^2 ,x_{Bj}  < 0.05)}^{CC\;TOT} }}} \right.
 \kern-\nulldelimiterspace} {\sigma _{(Q^2  > 200GeV^2 ,x_{Bj}  < 0.05)}^{CC\;TOT} }}
  = \left( {2.9 \pm 1.2\left( {stat} \right) \pm 0.8\left( {syst} \right)} \right)\% , 
 \end{array}
\]
is similar to that measured in the NC DIS process\cite{ZEUSNCDIFF} in a similar kinematic region.

\section{HERA-II Results}
\begin{figure}
\begin{center}
\begin{minipage}[c]{0.48\textwidth}
\psfig{figure=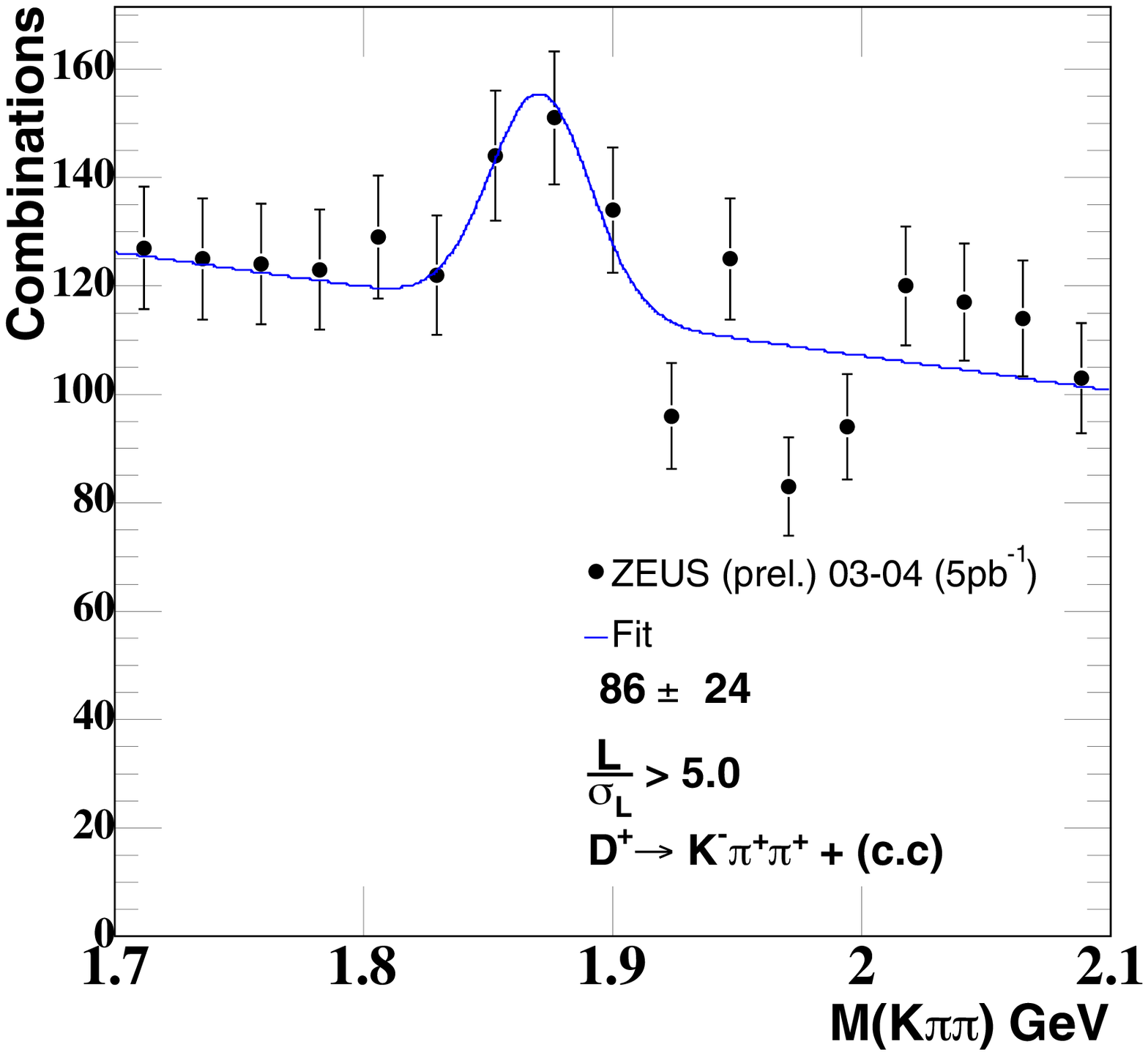, width={0.98\textwidth}}
\caption{$D^+$ signal extracted from HERA-II data using the new ZEUS microvertex detector.} 
\label{fig:dplus}
\end{minipage}
\hfill
\begin{minipage}[c]{0.48\textwidth}
\psfig{figure=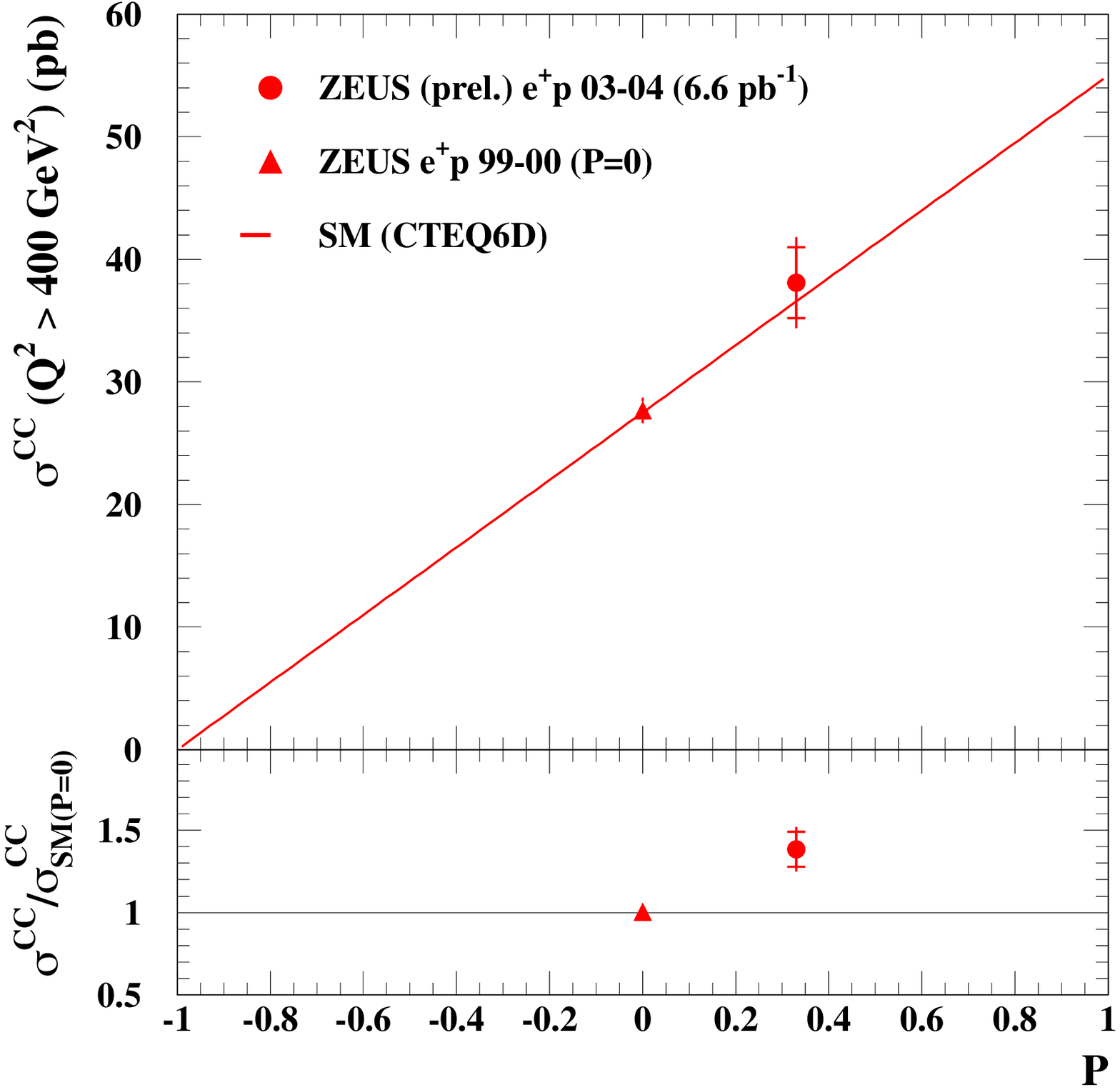, width={0.98\textwidth}}
\caption{Polarized and unpolarized CC DIS cross sections compared with
SM predictions.}
\label{fig:ccratio}
\end{minipage}
\end{center}
\end{figure}

Since the start of HERA-II running, ZEUS has been recording the
collisions of polarized positrons with protons using its new microvertex
and forward tracking detectors. Figure \ref{fig:dplus} shows use of the
new microvertex detector to place a significance cut on a secondary
vertex to extract a $D^+$ sample from 5 ${\rm pb}^{-1}$ of HERA-II data.  

Figure \ref{fig:ccratio} shows the ZEUS measurement of the polarized CC
DIS cross section for $Q^2 > 400 {\rm GeV}^2$ from 6.6 ${\rm pb}^{-1}$ of HERA-II
luminosity with $33\perc$ polarized positrons. The systematic error of
about $2\perc$ is principally due to the calorimeter energy scale, the
event selection, PDF uncertainty and the trigger acceptance. Also
plotted is the ZEUS HERA-I unpolarized point\cite{ZEUS} and the SM
prediction using the CTEQ6M\cite{Kretzer:2003it} PDF. Both data points
are shown to be consistent with the SM and the polarized positron CC
cross section has been measured, $\sigma_{CC} = 38.1 \pm 2.9 (stat.) \pm
(sys.) \pm 2.0 (lumi.) \pm (pol)$ pb.

\section{Summary and Conclusions}

New results from ZEUS are completing the picture from HERA-I. These
include the full complement of structure function and cross section
measurements. Precise jet measurements have determined $\alpha_s$ within
$2\perc$ and are being used to constrain the structure function QCD
fits. A new era in hadron spectroscopy has begun with the evidence of
pentaquark states. Charm physics is now providing new constraints on the
gluon distribution in the proton. DIS and photoproduction b cross
sections are now in general agreement with NLO QCD calculations. There
is also new understanding of diffraction in terms of QCD calculations. 

The first ZEUS results from HERA-II show great promise with new charm
data with the new ZEUS microvertex detector and the measurement of the
polarized CC DIS cross section. These provide just a taste of the rich
physics harvest to follow.

\begin{thebibliography}{0}

\bibitem{ijmp:a13:3385}A.M. Cooper-Sarkar, R.C.E. Devenish and
 A. De Roeck, Int. J. Mod. Phys. {\bf A 13},  3385 (1998). 
 
\bibitem{ZEUS} {ZEUS} Collab., S.~Chekanov {\it et al.}, ``High-$Q^2$
neutral current cross sections in $e^+ p$ deep inelastic scattering at
$\sqrt{s}~=~318$-GeV,'' arXiv:hep-ex/0401003.\\ {ZEUS} Collab.,
S.~Chekanov {\it et al.}, Eur.\ Phys.\ J.\ C {\bf 28}, 175 (2003).\\
{ZEUS} Collab., S.~Chekanov {\it et al.}, Eur.\ Phys.\ J.\ C {\bf 21},
443 (2001).\\ {ZEUS} Collab., J.~Breitweg {\it et al.}, Eur.\ Phys.\ J.\
C {\bf 11}, 427 (1999).
        
\bibitem{Kretzer:2003it} S.~Kretzer, H.~L.~Lai, F.~I.~Olness and
  W.~K.~Tung, JHEP {\bf 7}, 12 (2002).
  
\bibitem{dglap} V. N. Gribov and L. N. Lipatov, Sov. J. Nucl. Phys.
{\bf 15}, 438 (1972). \\ V. N. Gribov and L. N. Lipatov, Sov. J. Nucl.
Phys. {\bf 15}, 675 (1972).\\ Yu.L. Dokshitzer, Sov. Phys. JETP {\bf
46}, 641 (1977).\\ G. Altarelli and G. Parisi, Nucl. Phys. {\bf B 126},
298 (1977).

 \bibitem{Dokshitzer:1995zt}
Y.~L.~Dokshitzer and B.~R.~Webber,
Phys.\ Lett.\ B {\bf 352} (1995) 451.

\bibitem{Dasgupta:2001eq}
M.~Dasgupta and G.~P.~Salam,
Eur.\ Phys.\ J.\ C {\bf 24} (2002) 213.

\bibitem{NLOJET}
Z.~Nagy and Z. ~Trocsanyi, Phys.~Rev.~Lett. {\bf 87}, 082001 (2001).

\bibitem{MRST2001}
A.~D.~Martin \etal Eur.~Phys.~J. {\bf C23} (2002) 73.

\bibitem{LEPTO}
G.~Ingelman \etal Comp.~Phys.~Comm. {\bf 101}, 108 (1997).

\bibitem{FORWARD}
A.~H.~Mueller, Nucl.~Phys.~Proc.~Suppl. {\bf C18} 125 (1991).

\bibitem{ARIADNE}
L.~L\"onnblad, Comp.~Phys.~Comm.{\bf 71} 15 (1992);
L.~L\"onnblad, Z.~Phys. {\bf C 65}, 285 (1995).

\bibitem{DISENT}
S.~Catani and M.~H.~Seymour, Nucl.~Phys.~ {\bf B 485},
291 (1997); Erratum in Nucl.~Phys.~ {\bf B 510},
503 (1998).

\bibitem{zeuspq}
ZEUS \coll, S.~Chekanov \etal DESY-04-056 \mbox{hep-ex/0403051},
Phys.~Lett.~B (in press).

\bibitem{ks}
DIANA \coll, V.V.~Barmin \etal Phys.~Atom.~Nucl. {\bf 66} (2003) 1715;
A.E.~Asratyan, A.G.~Dolgolenko, M.A.~Kubantsev, Preprint
\mbox{hep-ex/0309042}, 2003; SVD \coll, A.~Aleev \etal Preprint
\mbox{hep-ex/0401024}, 2004; HERMES \coll, A.~Airapetian \etal
Phys.~Lett. {\bf B~585} (2004) 213; COSY-TOF \coll, M.~Abdel-Bary \etal
Preprint \mbox{hep-ex/0403011}, 2004.

\bibitem{pr:d67:012007}
ZEUS Coll., S. Chekanov {\it et al.}, Phys. Rev. {\bf D67} (2003) 012007.

\bibitem{pl:b528:199}
H1 Coll., C. Adloff {\it et al.}, Phys. Lett. {\bf B528} (2002) 199.

\bibitem{epj:c12:35}
ZEUS Coll., J. Breitweg {\it et al.}, Eur. Phys. J. {\bf C12} (2000) 35.

\bibitem{ZEUSelec}
ZEUS Coll., J. Breitweg {\it et al.}, Eur. Phys. J. {\bf C18} (2001) 625.

\bibitem{beautyphp}
ZEUS Coll., S. Chekanov {\it et al.}, DESY Preprint DESY-03-212, 2003.

\bibitem{fmnr}
S. Frixione {\it et al.}, Nucl. Phys. {\bf B412} (1994) 225.

\bibitem{HVQDIS}
B.W. Harris and J. Smith, Phys. Rev. {\bf D57} (1998) 2806.

\bibitem{epj:c24:345}
ZEUS Coll., S. Chekanov {\it et al.}, Eur.~Phys.~J. {\bf C 24} (2002) 345.

\bibitem{pr:d62:14022}
A.D. Martin, M.G. Ryskin and T. Teubner, Phys. Rev. {\bf D 62} (2000) 14022.

\bibitem{epj:c23:73}
A.D. Martin {\it et al.}, Eur.~Phys.~J. {\bf C 23} (2002) 73.

\bibitem{JHEP:0207:012}
J. Pumplin  {\it et al.}, JHEP {\bf 0207} (2002) 012.

\bibitem{EPJC5}
ZEUS Coll., S. Chekanov {\it et al.}, Eur~ Phys.~J. {\bf C 5} (1998) 41.

\bibitem{DJANGOH}
H. Spiesberger, www.desy.de/hspiesb/heracles.html (1998).

\bibitem{RAPGAP}
H. Jung, Comp. Phys. Comm. {\bf 86} (1995) 147.

\bibitem{ZEUSNCDIFF}
ZEUS Coll., J. Breitweg {\it et al.}, Eur. Phys. J. {\bf C 6} (1999) 43.

\end{thebibliography}
\end{document}